\documentclass[11pt]{article}
\usepackage{graphicx}
\usepackage{subfigure}
\usepackage{epsfig}
\usepackage{times}
\usepackage{lscape}
\usepackage{rotating}
\usepackage{longtable}
\usepackage{multirow}
\usepackage{xspace}
\usepackage{subfigure}
\usepackage{url}
\usepackage[numbers]{natbib}
\usepackage{epic}
\usepackage{overpic}

\newcommand{\BABARPubYear}    {08}

\newcommand{\BABARConfNumber} {005 }
\newcommand{\SLACPubNumber} {13300}

\newcommand{\twowidefig} {0.48}

\input pubboard/babarsym

\def\er #1 #2 { $#1 \pm #2$ }

\def\btn {\ensuremath {\B^+ \to \tau^+ \nu_{\tau}}\xspace}

\def\btodszlnu {\ensuremath {\B^- \to D^{*0} \ell^- \overline{\nu}_\ell}\xspace}

\def\tautoenunu {\ensuremath {\tau^+ \to e^+ \nu_e \overline{\nu}_\tau}\xspace}
\def\tautoe {\ensuremath {\tau^+ \to e^+ \nu_e \overline{\nu}_\tau}\xspace}

\def\tautomununu {\ensuremath {\tau^+ \to \mu^+ \nu_{\mu} \overline{\nu}_{\tau}}\xspace}
\def\tautomu {\ensuremath {\tau^+ \to \mu^+ \nu_{\mu} \overline{\nu}_{\tau}}\xspace}

\def\tautopinu {\ensuremath {\tau^+ \to \pi^+ \overline{\nu}_{\tau}}\xspace}
\def\tautopi {\ensuremath {\tau^+ \to \pi^+ \overline{\nu}_{\tau}}\xspace}

\def\tautopipiznu {\ensuremath {\tau^+ \to \pi^+ \pi^{0} \overline{\nu}_{\tau}}\xspace}
\def\tautopipiz {\ensuremath {\tau^+ \to \pi^+ \pi^{0} \overline{\nu}_{\tau}}\xspace}
\def\tautorho {\ensuremath {\tau^+ \to \pi^+ \pi^{0} \overline{\nu}_{\tau}}\xspace}

\def\tautothreepinu {\ensuremath {\tau^+ \to \pi^+ \pi^{-} \pi^{+} \overline{\nu}_{\tau}}\xspace}

\def\btodlnu      {\ensuremath{B^{-}\rightarrow\Dz\ell^{-}\bar{\nu} X}\xspace}

\def\btodordszlnu {\ensuremath{\Bub\to D^{(*)0}\ellm\bar{\nu}X}\xspace}

\def\eextra       {\ensuremath{E_{\mbox{\scriptsize{extra}}}}}

\def\NBG          {\ensuremath{N_{\mathrm{BG}}}}
\def\Nobs          {\ensuremath{N_{\mathrm{obs}}}}
\def\Nsig         {\ensuremath{N_{\mathrm{sig}}}}

\def\pprimesignal {\ensuremath{p^{'}_{{\mathrm{sig}\ \ell}}}}

\def\bln {\ensuremath {\B^+ \to \ell^+ \nu_{\ell}}\xspace}
\def\ben {\ensuremath {\B^+ \to e^+ \nu_{e}}\xspace}
\def\bmun{\ensuremath {\B^+ \to \mu^+ \nu_{\mu}}\xspace}

\def\lhrbb  {\ensuremath {LHR_{B \overline{B}}}}
\def\lhrcont{\ensuremath {LHR_{\mathrm{cont.}}}}

\def\nul90{\ensuremath{\mu_{s}^{90}}}

\def\nBB    {\ensuremath {(458.9 \pm 5.1) \times 10^{6}} }

\def\onlumi {\ensuremath { 417.6  \invfb\ }}

\def\offlumi   {\ensuremath { 42.2 \invfb\  }}

\def\btnresult {\ensuremath {\left(1.8 \pm 0.8 \pm 0.1  \right) \times 10^{-4}} }

\def\btnsig {\ensuremath{2.4 \sigma}}

\def\btntautoeresult {\ensuremath {\left(4.0 \pm 1.2  \right) \times 10^{-4}} }
\def\btntautomuresult {\ensuremath {\left(1.0^{+1.2}_{-0.9}  \right) \times 10^{-4}} }
\def\btntautopiresult {\ensuremath {\left(0.6^{+1.1}_{-0.5}  \right) \times 10^{-4}} }
\def\btntautorhoresult {\ensuremath {\left(2.0^{+1.4}_{-1.3}  \right) \times 10^{-4}} }

\def\btnresultcomb {\ensuremath {\left(1.8 \pm 0.6  \right) \times 10^{-4}} }

\def\benlimit  {\ensuremath {7.7 \times 10^{-6}} }
\def\bmunlimit  {\ensuremath {11 \times 10^{-6}} }
\def\btnlimit  {\ensuremath {3.2 \times 10^{-4}} }

\def\fB {\ensuremath {230 \pm 57 \mev} }

\long\def\inst#1{\par\nobreak\kern 4pt\nobreak
    {\it #1}\par\vskip 10pt plus 3pt minus 3pt}

\begin{document}
{\pagestyle{empty}

\begin{flushright}
\babar-CONF-\BABARPubYear/\BABARConfNumber \\
SLAC-PUB-\SLACPubNumber \\
September 2008 \\
\end{flushright}

\par\vskip 5cm

\begin{center}
\Large \bf A Search for {\boldmath $\bln$} \xspace Recoiling Against {\boldmath \btodlnu}
\end{center}
\bigskip

\begin{center}
\large The \babar\ Collaboration\\
\mbox{ }\\
\today
\end{center}
\bigskip \bigskip

\begin{center}
\large \bf Abstract
\end{center}
We present a search for the decay $\bln\ (\ell = \tau, \mu, \mathrm{or}\ e)$ in \nBB \xspace $\FourS$
decays recorded with the \babar\ detector at the SLAC PEP-II $B$-Factory. 
A sample of events with one reconstructed 
exclusive semi-leptonic $B$ decay (\btodlnu) is selected, and
in the recoil a search for $\bln$ signal is performed. The $\tau$ is
identified in the following channels: $\tautoenunu$, $\tautomununu$,
$\tautopinu$, and $\tautopipiznu$. The analysis strategy and 
the statistical procedure is set up for branching fraction 
extraction or upper limit determination. We determine from the dataset a preliminary measurement of 
$\mathcal{B}(\btn) = \btnresult$, which excludes zero at \btnsig, and $f_{B} = \fB$.  Combination with the hadronically tagged measurement yields $\mathcal{B}(\btn) = \btnresultcomb$. We also set preliminary limits on the branching fractions at  $\mathcal{B}(\ben)  < \benlimit\ (\textrm{90\% C.L.})$, $\mathcal{B}(\bmun)  < \bmunlimit\ (\textrm{90\% C.L.})$, and $\mathcal{B}(\btn)  < \btnlimit (\textrm{90\% C.L.})$.
\vfill
\begin{center}

Submitted to the 5$^{\rm th}$ International Workshop on the CKM Unitarity Triangle\\
9---13 September 2008, Rome, Italy.

\end{center}

\vspace{1.0cm}
\begin{center}
{\em Stanford Linear Accelerator Center, Stanford University, 
Stanford, CA 94309} \\ \vspace{0.1cm}\hrule\vspace{0.1cm}
Work supported in part by Department of Energy contract DE-AC02-76SF00515.
\end{center}

\newpage
}

\begin{center}
\small

The \babar\ Collaboration,
\bigskip

%
B.~Aubert,
M.~Bona,
Y.~Karyotakis,
J.~P.~Lees,
V.~Poireau,
E.~Prencipe,
X.~Prudent,
V.~Tisserand
\inst{Laboratoire de Physique des Particules, IN2P3/CNRS et Universit\'e de Savoie, F-74941 Annecy-Le-Vieux, France }
J.~Garra~Tico,
E.~Grauges
\inst{Universitat de Barcelona, Facultat de Fisica, Departament ECM, E-08028 Barcelona, Spain }
% ...

\end{center}\newpage

\section{INTRODUCTION}
\label{sec:Introduction}

\begin{figure}[htb]
\begin{center}
\includegraphics[width=0.45\textwidth]{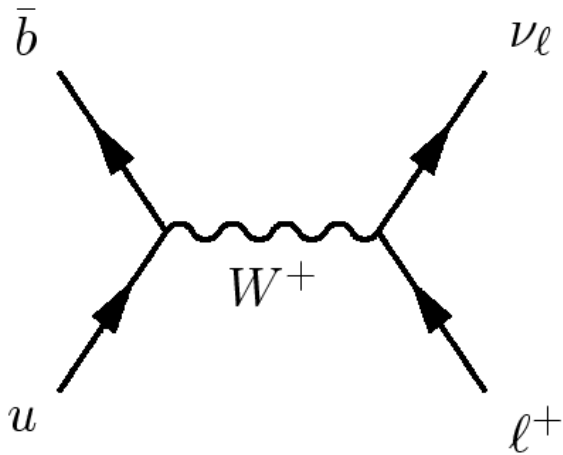}
\end{center}
\caption{\label{fig:feynman_diagram}%
The purely leptonic $B$ decay $\Bu \to \tau^{+} \nu_{\tau}$  
proceeding via quark annihilation into a $W^{+}$ boson.}
\end{figure}

In the Standard Model (SM), the purely leptonic decay \bln\ 
\footnote{Charge-conjugate modes are implied throughout this paper. The signal $B$ will be denoted as a \Bu\ decay while the semi-leptonic $B$ will be denoted as a \Bub.}
proceeds via quark annihilation 
into a $W^{+}$ boson (Fig. \ref{fig:feynman_diagram}).

The branching fraction is given by:
\begin{equation}
\label{eqn:br}
\mathcal{B}(\bln)= 
\frac{G_{F}^{2} m^{}_{B}  m_{\ell}^{2}}{8\pi}
\left[1 - \frac{m_{\ell}^{2}}{m_{B}^{2}}\right]^{2} 
\tau_{\Bu} f_{B}^{2} |V_{ub}|^{2},\label{eq:brsm} 
\end{equation}
where we have set $\hbar = c = 1$,
$G_F$ is the Fermi constant, 
$V_{ub}$ is a quark mixing matrix element~\cite{c,km}, 
$f_{B}$ is the $\Bu$ meson decay constant, which describes the
overlap of the quark wave-functions inside the meson,
$\tau_{\Bu}$ is the $\Bu$ lifetime, and
$m^{}_{B}$ and $m_{\tau}$ are the $\Bu$ meson and $\tau$ masses.
This expression is entirely analogous to that for pion decay.
Physics beyond the SM, such as two-Higgs doublet models,
could enhance or suppress the $\mathcal{B}(\bln)$ through the
introduction of a charged Higgs boson~\cite{HiggsPaper}.

Current theoretical values for $f_B$ 
(obtained from lattice QCD calculations)~\cite{UT2006} have  
large uncertainties, and purely leptonic decays of the $\Bu$ meson
may be the only clean experimental method of measuring
$f_B$ precisely. Given measurements of $|V_{ub}|$ from semi-leptonic
$B \to\u\ell\nu$ decays, $f_{B}$ could be extracted
from the measurement of the \btn\ branching fraction. In addition,
by combining the branching fraction measurement with results
from $B$ mixing, the ratio $|V_{ub}|/|V_{td}|$ can be extracted
from $\mathcal{B}(\btn)/\Delta m$, where $\Delta m$ is the
mass difference between the heavy and light neutral $B$ meson states.

The SM estimates of these branching fraction are 
$\BR(\btn) = (1.2 \pm 0.4)\times 10^{-4}$, 
$\BR(\bmun) = (5.6 \pm 1.7)\times 10^{-7}$, and 
$\BR(\ben) = (1.3 \pm 0.4)\times 10^{-11}$.  
We use $|V_{ub}| = (4.43 \pm 0.54)\times 10^{-3}$~\cite{vub} and a theoretical calculation of $f_{B}~=~189\pm 27$~MeV~\cite{UT2006} in Eq.~\ref{eq:brsm}.  The differences in the branching fractions are due to helicity supression, which is expressed in Equation \ref{eqn:br} via the different masses of the leptons.

\section{THE \babar\ DETECTOR AND DATASET}
\label{sec:babar}

The data used in this analysis were collected with the \babar\ detector
at the \pep2\ storage ring. 
The sample corresponds to an integrated
luminosity of \onlumi \xspace with center-of-mass energy equal to the  \FourS\ rest mass (on-resonance) 
and \offlumi \xspace taken $40\mev$ below $B\bar{B}$ threshold 
(off-resonance). The on-resonance sample consists of
about $\nBB$  $\FourS$ decays (\BB\ pairs). The collider is operated with asymmetric
beam energies, producing a boost of $\beta\gamma \approx 0.56$ 
of the \FourS\ along the collision axis.

The \babar\ detector is optimized for asymmetric energy collisions at a
center-of-mass (CM) energy corresponding to the \FourS\ resonance.
The detector is described in detail in Ref.~\cite{babar}. 
The components used in this analysis are the tracking system
composed of a five-layer silicon vertex detector and a 40-layer drift chamber, the Cherenkov detector for charged $\pi$--$K$ discrimination, the CsI calorimeter for photon and electron identification, and the flux return located 
outside of the 1.5T solenoidal coil and instrumented with resistive plate chambers for muon
and neutral hadron identification. For the most recent 121 \invfb\ of data, a portion of the muon 
system has been upgraded to limited streamer tubes~\cite{Benelli:2006pa,Menges:2006xk}.

A GEANT4-based \cite{geant4} Monte Carlo (MC) 
simulation is used to model the signal
efficiency and the physics backgrounds. Simulation samples
equivalent to approximately three times the accumulated data  were
used to model \BB\ events, and samples equivalent to approximately
1.5 times the accumulated data were used to model continuum events where
$\epem \to$ \uubar, \ddbar, \ssbar, \ccbar\ and \tautau. 
Three large samples of signal events are simulated, where
a $B^+$ meson decays to $e^+\nu_{e}$ ($7.8 \times 10^{6}$ events), $\mu^+\nu_{\mu}$ ($7.6 \times 10^{6}$ events), or $\tau^+\nu_{\tau}$ ($19.4 \times 10^{6}$ events), and a $B^-$  
meson decays to an acceptable $B$ mode. Beam related background and detector 
noise from data are overlaid on the simulated events.

\section{ANALYSIS METHOD}
\label{sec:Analysis}

Due to the presence of at least one neutrino in the final state, the \bln \xspace decay modes
lack the kinematic constraints that are usually exploited in $B$ decay 
searches to reject both continuum and $\BB$ backgrounds.
The strategy adopted for this analysis is to reconstruct exclusively 
the decay of one of the $B$ mesons in the event, referred to as the ``tag'' $B$. 
The remaining particle(s) in the event, referred to as the 
``signal $B$'', are then compared with the signatures
expected for \bln. In order to avoid experimenter bias, the 
signal region in data is not examined (``blinded'') until the final yield
extraction is performed.

The tag $B$ is reconstructed in the set of semileptonic $B$ decay modes 
\btodlnu, where $\ell$ is $e$ or $\mu$ and $X$ can be either nothing or a 
transition particle from a higher mass charm state decay, which we do not attempt 
to reconstruct  (although those tags consistent with neutral $B$ decays are vetoed).
The $\Dz$ is reconstructed in four decay modes:
$K^{-}\pi^{+}$, $K^{-}\pi^{+}\pi^{-}\pi^{+}$, $K^{-}\pi^{+}\pi^{0}$, and
$K_{s}^{0}\pi^{+}\pi^{-}$. The $K_{s}^{0}$ is reconstructed only in the
mode $K_{s}^{0} \rightarrow \pi^{+}\pi^{-}$.
In a previous search for $\btn$ \cite{babar_sl_btn} we found that the low momentum transition daughter of $D^{*0}$ decays need not be reconstructed.   Reconstructing the final state $B\to\Dz\ell\nu X$ provides a higher efficiency but
somewhat lower purity than the exclusive reconstruction method of $\btodszlnu$.  In this analysis, we employ a technique, known as the ``seeding'' method, to recapture one photon from the $X$ state.   The decays $\ben$ and $\bmun$ have not been previously searched for in the recoil of semileptonic tags.

Since the $\tau$ decays before reaching active detector elements, the \btn \xspace signal is searched for in
both leptonic and hadronic $\tau$ decay modes:
$\tautoenunu$, $\tautomununu$, $\tautopinu$, and $\tautopipiznu$. The branching fractions of the above $\tau$ decay 
modes are listed in Table \ref{tab:TauDecayModes}.

\begin{table}[h]
\caption{\label{tab:TauDecayModes} Branching fractions for the $\tau$ decay modes used in the \btn\ search~\cite{ref:pdg2006}.}
\begin{center}
\begin{tabular}{|l|c|}
\hline
Decay Mode   &   Branching Fraction (\%)  \\
\hline
$\tautoenunu$      & 17.84 $\pm$ 0.05 \\ 
$\tautomununu$     & 17.36 $\pm$ 0.05 \\ 
$\tautopinu$       & 10.90 $\pm$ 0.07 \\ 
$\tautopipiznu$    & 25.50 $\pm$ 0.10 \\ 
\hline
\end{tabular}
\end{center}
\end{table}

\subsection{Tag \boldmath{B} Reconstruction}
\label{sec:TagReco}

The tag $B$ reconstruction proceeds as follows. First, we reconstruct the 
$\Dz$ candidates in the aforementioned four decay modes using reconstructed tracks
and photons where a $\piz$ is included. The tracks
are required to meet particle identification criteria consistent
with the particle hypothesis, and are required to converge at a common vertex.
The $\piz$ candidate is required to have invariant mass between 
0.115--0.150 \gevcc\ and its daughter photon candidates must 
have a minimum energy of 30 \mev.
The mass of the reconstructed $\Dz$ candidates in 
$K^{-}\pi^{+}$, $K^{-}\pi^{+}\pi^{-}\pi^{+}$, and $K_{s}^{0}\pi^{+}\pi^{-}$
modes are required to be within 20 \mevcc\ of the nominal mass 
\cite{ref:pdg2006}.
In the $K^{-}\pi^{+}\pi^{0}$ decay mode, 
the mass is required to be within 35 \mevcc\ of the nominal mass
\cite{ref:pdg2006}; this wider mass window accounts for the \babar\ detector' lower mass resolution when reconstructing particle candidates from neutral clusters, as opposed to reconstructing candidates involving charged tracks.

Finally, $\Dz\ell$ candidates are reconstructed by combining the
$\Dz$ with an identified electron 
or muon with momentum above 0.8 \gevc\ in the CM frame. 
The $D^{0}$ and $\ell$ candidates are required to meet at a common vertex.  If more than one suitable $\Dz\ell$ candidate is 
reconstructed in an event, the best candidate is taken to be the
one with the highest vertex probability. 
The uncorrected
tag reconstruction efficiency in the signal MC simulation is 
1.7\% for $\btn$,
1.1\% for $\bmun$ and
1.1\% for $\ben$.

\subsection{Selection of \boldmath{\bln} \xspace signal candidates}
\label{sec:SigSelection}

After the tag $B$ reconstruction, in the signal $B$, we identify one of the following reconstructed particles: $e^{+}$, $\mu^{+}$, $\pi^{+}$, or $\rho^{+}$.  The  $e^{+}$ and $\mu^{+}$ can come from  $\btn$ or directly from $\bmun$ or $\ben$.  Each signal $B$ track must satisfy the following selection criteria: its point of closest approach to the interaction point is 
less than 2.5~\cm\ along the beam axis and less than 1.5~\cm\ transverse 
to the beam axis.

The different reconstructed particles are assigned using a hierarchical selection involving kinematic constraints and particle identification.  All of the signal decay modes for which we search contain only one track from the signal $B$.  If more tracks that match the criteria stated above are present after a tag $B$ has been reconstructed, the event is rejected. If the track from the signal $B$ is identified as a kaon, it is rejected.  Since we search for only one track in the signal $B$, we classify that track as one of the following in the priority given.

\begin{itemize}

\item If the track satisfies particle identification as a muon, it is classified as such.

\item If the track satisfies particle identification as an electron, it is classified as such.  We apply Bremsstrahlung radiation recovery techniques to identify as many electrons as possible.

\item If the track can be combined with a $\pi^{0}$ to form a $\rho^{+}$ with a common vertex, it is classified as such.  The invariant mass of a $\piz$ candidate
must be between 0.115--0.150 \gevc;
the shower shape of the daughter photon candidates must be consistent with 
an electromagnetic shower shape, and the photons
must have a minimum energy of 50 \mev\ in the CM frame. 

\item If the track is not accepted by any of the above filters, it is classified as a $\pi^{+}$ by default.

\end{itemize}

\noindent Background consists primarily of $B^{+}B^{-}$ events in which the tag
$B$ meson has been correctly reconstructed and the recoil side contains
one signal candidate track and additional particles which are not 
reconstructed by the tracking detectors or calorimeters. Typically these
events contain  $K_{L}^{0}$ candidates and/or neutrinos, and frequently
also additional charged or neutral particles which pass outside of the 
tracking and calorimeter acceptance. Background events also contain
$B^{0}\bar{B}^{0}$ events. In addition some excess events in data,
most likely from two-photon and QED processes which are not modeled in the MC 
simulation, are also seen. 

Multiple variables are used to suppress backgrounds.  Most are combined into two likelihood ratios (LHRs), which are probability distributions designed to produce maximum separation between signal and background.  Two variables are reserved for individual use due to their discriminating power.  They are the momentum of a signal lepton in the $B^{+}$ rest frame ($\pprimesignal$) and the total energy recorded in the detector that is not assigned to the tag or signal $B$ ($\eextra$).

Due to the presence of the neutrino in the products of the tag $B$, the direction of neither $B$ can be known accurately.
Instead, $\cos \theta_{B-D^{0} \ell}$ (the cosine of the angle between the $\Dz\ell$ candidate and the $B$ meson momenta) is calculated in the $\Upsilon(4S)$ rest frame. 

\begin{equation}
\label{eq:CosBY}
\cos\theta_{B-D^{0}\ell} = \frac{2 E_{B} E_{D^{0}\ell} - m_{B}^{2} - m_{D^{0} \ell}^{2}}{2|\vec{p}_{B}||\vec{p}_{D^{0}\ell}|},
\end{equation}
where ($E_{D^{0}\ell}$, $\vec{p}_{D^{0}\ell}$) and
($E_{B}$, $\vec{p}_{B}$) are the 
four-momenta in the CM frame, and $m_{D^{0}\ell}$ and $m_{B}$ 
are the masses of the $D^{0}\ell$ candidate and tag $\Bm$ meson, respectively. 
$E_{B}$ and the magnitude of $\vec{p}_{B}$ are calculated 
from the beam energy: $E_{B} = E_{\rm{CM}}/2$ and 
$ | \vec{p}_{B} | = \sqrt{E_{B}^{2} - m_{B}^{2} }$.
This definition assumes that the only missing particle in the tag $B$ decay
is a massless neutrino.
Events in which the tag $B$ daughters include a $D^{0}$ and no higher mass charmed states are more common in the physical region, but other events have a larger tail into the non-physical region $\cos\theta_{B-D^{0}\ell} < -1$.

For the reconstructed leptons in the signal $B$, we estimate the momentum of the signal lepton in the signal $B^{+}$ rest frame ($\pprimesignal$) by averaging around the cone formed by $\cos\theta_{B-D^{0}\ell}$.  Since $\ben$ and $\bmun$ are two-body decays, for true signal events, $\pprimesignal$ should exhibit a peak at 
\begin{equation}
\pprimesignal = \frac{m_{B}^{2} - m_{\ell}^{2}}{2m_{B}} \approx \frac{m_{B}}{2} = 2.64 \gevc.  
\end{equation}

If an event has a reconstructed signal muon candidate and $\pprimesignal > 2.3 \gevc$, it is classified as a $\bmun$ candidate; otherwise it is classified as a $\tautomununu$ candidate.  If an event has a reconstructed signal electron candidate and $\pprimesignal > 2.25 \gevc$, it is classified as a $\ben$ candidate; otherwise it is classified as a $\tautoenunu$ candidate.

In an ideal $\bln$ decay, we reconstruct all tracks and clusters associated with the real decay.  The only unreconstructed particles would be neutrinos, which leave no energy in the detector.  Therefore, we expect our signal to concentrate near zero $\eextra$. We require a minimum energy of $30 \mev$ for any neutral cluster.

After the $D^{0}$ has been reconstructed, a ``seeding'' algorithm adds a photon (called the ``seed photon'') to the reconstructed $D^{0}$ and reevaluates $\cos\theta_{B-D^{0}\ell}$.  The seeding algorithm performs this procedure with all photons that do not overlap with the tag $B$ and have CM energy less than 300 \mev. 
If a seed photon causes $\cos\theta_{B-D^{0}\ell}$ to become closer to (but not greater than) 1, it is selected.  We seek to move $\cos\theta_{B-D^{0}\ell}$ closer to 1 because events containing real $D^{*}$ mesons usually appear in the low tail of the $\cos\theta_{B-D^{0}\ell}$ distribution.  If more than one photon satisfies these conditions, the one which moves $\Delta M \equiv m_{\Dstarz} - m_{\Dz}$ closest to the nominal value of 142.12 \mevcc \cite{ref:pdg2006} is used.  The photon is removed from $\eextra$ and the event is stored with the modified $\eextra$ and $\cos\theta_{B-D^{0}\ell}$ variables.  

We use a single photon to account for decays such as $D^{*0} \to \Dz \gamma$ and $D^{*0} \to \Dz \piz\ (\piz \to \gamma \gamma)$.  We studied the possibility of including a second seed photon, but it did not produce a significant improvement in performance.

\subsection{Likelihood Ratios}
\label{sec:LHRs}

To take advantage of shape differences between variables, we use two LHRs that consist of several probability density functions (PDFs).  Separate LHRs are generated for $B\bar{B}$ and continuum background suppression.  Two PDFs are generated for each variable in a LHR. One uses the 
signal MC sample and is treated as a probability $P_{s} (x)$, where $x$ is the value of the PDF variable.  The other uses the relevant generic MC samples and is treated as a probability $P_{b}(x)$.  These two PDFs are combined to form a probability $P_{i} (x)$
\begin{equation}
P_{i} (x) = \frac{P_{s} (x)}
                 {P_{s} (x) + P_{b}(x)},
\end{equation}
\noindent where each $i$ represents a different variable.  Bins in a distribution that are more likely to contain background events have  $P_{i} (x)$ closer to 0; bins that are more likely to contain signal events have  $P_{i} (x)$ closer to 1.  Each LHR is formed by multiplying all $P_{i} (x)$ together:

\begin{equation}
{\rm LHR}(x) \equiv \prod_{i} P_{i} (x).
\end{equation}

Ideally, a LHR is a doubly peaked distribution with background events forming a peak near zero and signal events forming a peak near 1.  The PDFs are created using MC samples with all tag selection criteria applied.  Any given event will have one LHR for $B\bar{B}$ events and one for continuum events, where the PDFs are selected based on the reconstructed decay mode of that event.

\subsubsection{Variables Included in LHR}
\label{sec:LHRvars}

Multiple variables were considered for inclusion in the LHRs.  For each of the 14 LHRs (7 decay modes $\times$ 2 background types), a fixed signal yield was chosen.  Each LHR was tested using signal MC samples and the background MC samples it was designed to reject.  The test were performed on MC after the appropriate decay mode was selected and with $\eextra$ required to be less than $1.5 \gev$.  For a given LHR, a baseline performance was calculated by using all prospective variables.  A cut was placed on the LHR in question to produce the chosen signal yield, and the Punzi Figure of Merit (${\rm FOM_{Punzi}}$) was calculated \cite{Punzi:2003bu}.

\begin{equation}
\label{eq:punzi}
{\rm FOM_{Punzi}} = \frac{\Nsig}{N_{\sigma}/2 + \sqrt{\NBG}},
\end{equation}
\noindent where $\Nsig$ is the signal yield and $\NBG$ is the background yield.  $N_{\sigma}$ is the number of standard deviations desired from the result.  We use   $N_{\sigma} = 3$.

The Punzi FOM is better suited to searches for small signals on small backgrounds.  It is designed to prevent optimization algorithms from reducing the background to zero and creating a undesirably low signal yield.  We used it for the LHRs for all modes because significance ($\Nsig/\sqrt{(\Nsig+\NBG)}$) sometimes caused our optimization algorithm to reduce the background to zero.

Each PDF was tested by removing it (and {\it only} it) from the LHR.  The LHR was scanned again until the chosen signal yield was reached, and the FOM was recalculated.  If removing the PDF increased the FOM by a statistically significant amount, it was not included in the final LHR.  We also removed any variables that should not, for physical reasons, improve the analysis.  For instance, variables related to the tag $B$ would not reject $B\bar{B}$ background because the vast majority of it has a properly reconstructed tag $B$. The results of this selection process are shown in Table \ref{tab:PDFsUsed}.  Four of these variables were combined to form 2-D PDFs.  The remainder were used as 1-D PDFs. 

\begin{itemize}

\item{Separation Between the Signal and Tag $B$ Vertices ($\Delta z / \sigma_{\Delta z}$):}

Due to the neutrinos on both sides of the event, the vertices of the reconstructed signal tracks and neutral clusters do not correspond exactly to the true $B$ decay points.  However, the reconstructed vertices are still displaced in space while tracks from continuum processes tend to point back to the interaction point.  

We calculate the displacement between the putative $B$ vertices divided by the uncertainty on that displacement.  Continuum events are distributed more strongly towards zero than true $B\bar{B}$ events.

\item{Net Event Charge:}

In our previous search \cite{babar_sl_btn}, we noted a drop in tag efficiency over the lifetime of the experiment.  We found that approximately half of this drop was due to a requirement that the event have zero net charge in order to pass tag selection.  To avoid this drop in efficiency while retaining the discriminating power of this variable, the net charge of the event was moved from tag selection to the LHRs.

\item{Ratio of the Second to Zeroth Fox-Wolfram Moment ($R2_{\rm All}$):}

The Fox-Wolfram moments are rotationally invariant kinematic quantities designed to quantify the shapes of events resulting from $e^{+}e^{-}$ collisions.   They are denoted $H_{l}$, where $l$ is the number of the moment.  
\begin{equation}
	H_{l} \equiv \frac{4 \pi}{2l + 1} \sum_{m = -l}^{+l} 
\left|
\sum_{i}Y_{l}^{m}(\theta_{i}) \frac{\vec{p_{i}}}{\sqrt{s}}
\right|^{2},
\end{equation}
where $i$ runs over all hadrons in the event, $\vec{p_{i}}$ are the momenta of the hadrons (in the CM frame), $Y_{l}^{m}$ are the spherical harmonics, $\theta$ is the angle of the momentum with respect to the $z$ axis, and $\sqrt{s}$ center-of-mass energy of the collision \cite{r2definition}.  This variable is the ratio 
\begin{equation}
	R2_{\rm All} \equiv \frac{H_{2}}{H_{0}}.
\end{equation}

We do not place a cut on this variable.  We use it as a PDF in the LHRs of several modes, just as we use all of the other variables in this list.

\item{$\cos\theta_{B-D^{0}\ell}$ : }

This angular variable is defined in Section \ref{sec:SigSelection} and Equation \ref{eq:CosBY}.  

\item{$D^0$ Decay Mode:}

Each $D^{0}$ decay mode is assigned an integer, and these integers form a distribution for an MC sample.  Signal and background events have different distributions of $\Dz$ decay mode.  For instance, true signal events are found recoiling against a $\Dz \to \Km \pip$ more often than $B \bar{B}$ background events.

\item{Center of Mass Momentum for the Tag $K^{-}$ ($p^{*}_{{\rm Tag} K^{-}}$) and $\ell^{-}$ ($p^{*}_{{\rm Tag} \ell^{-}}$)}

\item{Tag $K^{-}$ Selector:}

If the tag $\Dz$ decay produces a putative charged $K$, we assign an integer value to the track corresponding to level of $K$ particle identification passed by the track, with increasing values indicating tighter selection criteria.  This is an integer value ranging from 10-14; each track is assigned to exactly one of these numbers. The value 16 is assigned to $K_{s}^{0}$ candidates. Signal events concentrate more strongly than background events at higher values.

\item{Minimum Invariant Mass of Any 2 Reconstructed Tracks ($M_{2}^{\rm min}$):}

Since the minimum invariant mass of any three tracks $M_{3}^{\rm min}$ was a useful variable in our previous search \cite{babar_sl_btn}, we decide to try using $M_{2}^{\rm min}$.  As the name suggests, it is the smallest invariant mass produced by any combination of two tracks used to reconstruct the signal $B$.

\item{$m_{\ell \ell}$:}

If an event contains two putative lepton tracks, their invariant mass is calculated and stored as $m_{\ell \ell}$.  This variable was originally developed to remove pair-produced leptons.  It is obviously highly correlated with $M_{2}^{\rm min}$, so both variables are never used in the same LHR.

\item{Signal $\mu$ Selector:}

If the track from the signal $B$ decay passes the particle identification requirements to become a putative $\mu^{+}$, we check if it passes a stricter level of $\mu^{+}$ identification.  If so, the event is assigned the value one for this PDF.  Otherwise, it is assigned the value zero. More Continuum background than signal accumulates at zero.

\item{Signal $K^{+}$ Selector:}

We wish to suppress the misreconstruction of  $K^{+}$ as pions or leptons from the signal $B$, so we include $K$ selection in the LHR.  Specifically, this PDF is set to zero if a signal track passes loose $K^{+}$ particle identification requirements. 

\item{Vertex Status:}

For those $\tau$ modes that involve neutral clusters (i.e $\tautorho$), a vertex is created.  The quality of that vertex is reported as an integer from zero through four, which is included as  PDF.  Zero indicates that the vertex fit was successful.  Other values indicate various failure modes for the fit. The only failure mode that occurs in this analysis is that the fit does not converge

\item{ Reconstructed Mass of the $\tau$ Daughter ($m_{\rho^{+}}$):}

The decay $\tautopipiznu$ often proceeds through the $\rho^{+}$ resonance.  For true signal events, a peak at the resonance mass appears in the invariant mass distribution of the signal track and neutrals.  Background events yield a flat or linear distribution.  

\item{Center of Mass Momentum for 
the $\pip$ and $\piz$ in $\tautopipiznu$ ($p^{*}_{\pip}$, $p^{*}_{\piz}$).
}

\end{itemize}

\label{sec:PDF2D}

2-D PDFs are two-dimensional histograms that contain two variable distributions.  We use 2-D PDFs in cases where we want to exploit two variables that are highly correlated or that have a stronger separation when combined than when separate.

\begin{itemize}

\item{ $m_{\rm unreco} {\rm\ vs.\ } \cos(p_{\theta}^{\rm miss})$ ($m_{\rm unreco}$-dir):}

The total invariant mass and initial momentum of each event are well known from beam information.  Since neutrinos escape undetected from each event, we expect the total reconstructed invariant mass to be less than what the beam provides.  The difference is called the unreconstructed mass ($m_{\rm unreco}$).  The missing momentum $(p_{\theta}^{\rm miss})$ is similarly defined, where $\theta$ is the angle with respect to the beam line.

The PEP-II beam pipe corresponds to values of $\cos(p_{\theta}^{\rm miss})$ near $\pm 1$.  One source of background is events in which real particles are lost down the beam pipe, which is outside of the detector coverage.  Since they are not reconstructed, they can be misinterpreted as neutrinos.  This PDF allows the LHR to account for background events that have high unreconstructed mass but are likely to have lost particles down the beam pipe.

\item{$\cos\theta^{\prime}_{\tau-Y}{\rm\ vs.\ } |\vec{p^{\prime}}_{Y}|$ (Cos$\tau$Y-pY):}

$\cos\theta^{\prime}_{\tau-Y}$ is the equivalent of $\cos\theta_{B-D^{0} \ell}$ for the signal $B$.  $Y$ represents all of the reconstructed daughters of the signal $B$, and $\theta^{\prime}_{\tau-Y}$ is the calculated angle between $Y$ and $\tau$ momenta in the signal $B$ rest frame.   $|\vec{p^{\prime}}_{Y}|$ is the estimated momentum of $Y$ in the signal $B$ rest frame. It is the same as $\pprimesignal$ defined to include the hadronic $\tau$ decay modes; it is calculated with the same average around the cone formed by $\cos\theta_{B-D^{0} \ell}$.  Since $\pprimesignal$ is such a powerful variable for selected $\ben$ and $\bmun$, we tested it for $\btn$.  We found that it was not very useful unless combined in this PDF.  This PDF is not used in $\ben$ or $\bmun$ reconstruction because $\pprimesignal$ is kept as a separate variable. 

\end{itemize}

\begin{table}
\caption{This is a list of all variables used as signal PDFs for the various decay modes.  Variables in {\bf bold} are used for both Continuum and $B\bar{B}$ background. 
\label{tab:PDFsUsed}}
\footnotesize
\begin{center}
\begin{tabular}{|c|c|c|c|c|c|}
\hline
$\tautoe$ & $\tautomu$ & $\tautopinu$ & $\tautopipiz$  & $\bmun$ & $\ben$ \\
\hline
\hline
\multicolumn{6}{|c|}{\bf Net Charge} \\
\multicolumn{6}{|c|}{\boldmath $R2_{\rm all}$} \\
\multicolumn{6}{|c|}{$p^{*}_{{\rm Tag} \ell}$} \\
\multicolumn{6}{|c|}{ Tag $K^{-}$ Sel. Level} \\
\multicolumn{6}{|c|}{$\Dz$  Dec. Mode} \\ 
\hline
\multicolumn{4}{|c|}{$p^{*}_{{\rm Tag} K^{-}}$}& - &$p^{*}_{{\rm Tag} K^{-}}$ \\
\multicolumn{4}{|c|}{$\cos\theta_{B-D^{0}\ell}$} & - & - \\
\hline
{\boldmath $\Delta z/ \sigma_{\Delta z}$} & {\boldmath $\Delta z/ \sigma_{\Delta z}$} & {\boldmath $\Delta z/ \sigma_{\Delta z}$} & $\Delta z/ \sigma_{\Delta z}$ &  $\Delta z/ \sigma_{\Delta z}$ & {\boldmath $\Delta z/ \sigma_{\Delta z}$}\\
{\boldmath $m_{\ell \ell}$} & {\boldmath $m_{2}^{\rm min}$} & - & {\boldmath $m_{2}^{\rm min}$} & $m_{\ell \ell}$ &{\boldmath $m_{\ell \ell}$}  \\
{\boldmath \bf $m_{\rm unreco}$-dir} & {\boldmath \bf $m_{\rm unreco}$-dir} & - & - & {\boldmath \bf $m_{\rm unreco}$-dir} & {\boldmath \bf $m_{\rm unreco}$-dir} \\
 - & Signal $K^{+}$ Sel. & - & {\bf \boldmath Signal $K^{+}$ Sel.} &  - & - \\
 - & Signal $\mu$ Sel. & - & -  & {\bf \boldmath Signal $\mu$ Sel.} & - \\
Cos$\tau$Y-pY & {\bf \boldmath Cos$\tau$Y-pY} & {\bf \boldmath Cos$\tau$Y-pY} & - & - & - \\
 - & - & - & {\boldmath $p^{*}_{\pi^{+}}$}  & - & - \\
 - & - & - & {\boldmath $p^{*}_{\pi^{0}}$}  &   - & -\\
 - & - & - & {\boldmath $m_{\rho^{+}}$}  & -&- \\ 
 - & - & - & {\bf Vertex Status}  &- &- \\ 
\hline
\end{tabular}
\end{center}
\end{table}

\subsection{Optimization}
\label{sec:cutoptimization}

We use three ($\eextra$, $\lhrbb$, and $\lhrcont$) variables in our final selection of the five $\tau$ decay modes.  For $\ben$ and $\bmun$, we add a fourth variable $\pprimesignal$. The final requirements for these variables are obtained by optimizing on a figure of merit (FOM). For the $\tau$ decay modes, we choose the FOM to be significance ($\Nsig/\sqrt{(\Nsig+\NBG)}$).  For the other two leptonic $B$ decay modes, we use the Punzi FOM (Equation \ref{eq:punzi}).  

In the $\btn$ mode, both signal and background are large enough that optimizations perform well with standard significance.  We make this determination based on the $\Nsig$ expected from the SM predictions for the branching fractions.  The optimized selection criteria are shown in Table \ref{tab:optimizedcuts}.

\begin{table}[htb]
\begin{center}
\caption{Optimized ranges from which we accept signal candidates, which are mostly given by our optimization procedure.  The exceptions are the $\pprimesignal$ ranges for the two leptonic $\tau$ decay modes, which were chosen to separate the leptonic $\tau$ decays from  $\ben$ and $\bmun$.
\label{tab:optimizedcuts}
}
\begin{tabular}{|l|c|c|c|c|} \hline
Mode	&	$\eextra$ 	&	$\lhrbb$	&	$\lhrcont$	&	$\pprimesignal$	\\ \hline
$\tautoenunu$	&	[0,0.24] $\gev$	&	[0.74,1]	&	[0.16,1]	&	[0.00,2.25] \gevc	\\
$\tautomununu$	&	[0,0.24] $\gev$	&	[0.14,1]	&	[0.72,1]	&	[0.00,2.30] \gevc	\\
$\tautopinu$	&	[0,0.35] $\gev$	&	[0.57,1]	&	[0.80,1]	&	-	\\
$\tautopipiznu$	&	[0,0.24] $\gev$	&	[0.97,1]	&	[0.95,1]	&	-	\\ \hline
$\bmun$	&	[0,0.72] $\gev$	&	[0.33,1]	&	[0.75,1]	&	[2.45,2.92] \gevc	\\ \hline
$\ben$	&	[0,0.57] $\gev$	&	[0.00,1]	&	[0.01,1]	&	[2.52,3.02] \gevc	\\ \hline
\end{tabular}
\end{center}
\end{table}

After we examined the data in the signal region, we discovered an excess of data above our MC simulations at low values of $m_{\ell \ell}$, which is the minimum invariant mass of any two leptons.  These events constitute an unmodeled background and are most likely due to photon pair conversion in the material of the detector.  We decided remove all events below a certain value of $m_{\ell \ell}$ after all other analysis cuts had been applied.  This value was chosen using only signal and background MC simulations with the optimization technique described in this section.  The result excludes events in the $\tautoe$ channel with $m_{\ell \ell}<0.29 \gevcc$, which constitutes less than two percent of our signal MC sample.  All efficiencies and yields in this note include the effects of this requirement.

\subsection{Signal Efficiency}
\label{sec:SigEff}

The signal $B$ selection efficiencies 
for the  decay modes are 
determined from signal MC simulation and summarized 
in Table~\ref{tab:overalleff}.
The signal efficiencies correspond to the number of events
selected in a specific signal decay mode, given that a tag $B$ has
been reconstructed.

\begin{table}[htb]
\caption{Overall efficiency $(\eps \equiv \eps_{\rm sig} \times \eps_{\rm tag})$ of optimized signal selection for all modes before systematic corrections.}
\label{tab:overalleff} 
\begin{center} 
\begin{tabular}{|l|c|c|c|} \hline
Mode &	$\varepsilon_{\mathrm {sig}} $ & $\varepsilon (\times 10^{-4})$ \\ \hline
$\tautoenunu$	&	$(	1.987	\pm	0.043	) \% $&	3.38	$\pm$	0.07	\\
$\tautomununu$	&	$(	1.610	\pm	0.038	) \% $&	2.73	$\pm$	0.06	\\
$\tautopinu$	&	$(	2.48	\pm	0.05	) \% $&	4.21	$\pm$	0.08	\\
$\tautopipiznu$	&	$(	0.859	\pm	0.028	) \% $&	1.46	$\pm$	0.05	\\ \hline
$\btn$		&	$(	6.94	\pm	0.08	) \% $&	11.78	$\pm$	0.13	\\ \hline
$\bmun$		&	$(	30.92	\pm	0.36	) \% $&	32.54	$\pm$	0.36	\\ \hline
$\ben$		&	$(	36.98	\pm	0.38	) \% $&	40.43	$\pm$	0.40	\\ \hline
\end{tabular}
\end{center}
\end{table}

The selection efficiency for $\tautomununu$ is low compared to that of the
$\tautoenunu$ mode because the momentum spectrum 
of the signal muons peaks below 1.2 \gevc, where the muon detection
efficiency is low. Since no minimum momentum requirement and no tight pion
identification criteria are applied to the
$\tautopinu$ signal selection, electron and muon signal tracks 
that fail particle identification requirement get selected in this mode. 
Any true $\tautopipiznu$ signal events, with a missed $\piz$ 
are also included in $\tautopinu$ selection mode. 
Therefore, the $\tautopinu$ selection mode has the highest signal efficiency.

\subsection{Background Estimation from \boldmath{\eextra} Sidebands}
\label{sec:EextraSBExtrapolation}

We define the ``sideband'' region as $\eextra \ge 0.6 \gev$, except for $\bmun$ where it is defined as  $\eextra \ge 0.72 \gev$.
The ``signal region'' is defined separately for each signal mode using the optimized cuts on $\eextra$ given in Table \ref{tab:optimizedcuts}.  Distributions of $\eextra$ for the signal decay modes are shown in Figures \ref{fig:eextra_allcuts_unblind_e} - \ref{fig:eextra_allcuts_enu}.

For each mode, after applying the optimized final selections (except $\eextra$), 
the number of MC events in  the signal region ($N_{\mbox{\scriptsize{MC,Sig}}}$) 
and side band ($N_{\mbox{\scriptsize{MC,SideB}}}$) are counted
and their ratio ($R^{\mbox{\scriptsize{MC}}}$) is obtained. 

\begin{eqnarray*}
R^{\mbox{\scriptsize{MC}}} & = & \frac{N_{\mbox{\scriptsize{MC,Sig}}}}{N_{\mbox{\scriptsize{MC,SideB}}}}
\end{eqnarray*}

\noindent Using the number of data events in the side band ($N_{\mbox{\scriptsize{data,SideB}}}$)
and the ratio $R^{\mbox{\scriptsize{MC}}}$, the number of expected background events in the 
signal region in data ($N_{\mbox{\scriptsize{exp,Sig}}}$) is estimated. 

\begin{eqnarray*}
N_{\mbox{\scriptsize{exp,Sig}}} & = & N_{\mbox{\scriptsize{data,SideB}}} \cdot R^{\mbox{\scriptsize{MC}}}
\end{eqnarray*}

Table \ref{tab:BGpred_EEx} shows the background predictions from the $\eextra$ sideband.  We verify that the background predictions given by this sideband are consistent with the $D^{0}$ mass, $\lhrcont$, $\lhrbb$, and $\pprimesignal$ sidebands, where applicable. We also studied the predictions given when the $\eextra$ sideband is loosened to be $\geq 0.8 \gev$ or tightened to include all events but the signal region in each mode.  These predictions are also consistent with Table \ref{tab:BGpred_EEx}.

\begin{table}
\begin{center}
\caption{BG Predictions from $\eextra$ sideband.  
$R_{\rm MC}$ is the ratio of events in the sideband to events in the signal region of $\eextra$ in the background MC.
$N_{\rm data,SideB}$ is the number of events in the $\eextra$ sideband in data.
$N_{\rm MC,Sig}$ is the number of normalized events in the $\eextra$ signal region of the background MC samples.  This is the background prediction taken solely from the MC samples.
 $N_{\rm exp,Sig}$ is the product of $R_{\rm MC}$ and $N_{\rm data,SideB}$; it is the background prediction extrapolated from the data sideband using the MC samples.
}
\label{tab:BGpred_EEx}
\begin{tabular}{|l|c|c|c|c|}\hline
Mode & $R_{\rm MC}$ &	 $N_{\rm data,SideB}$ & $N_{\rm exp,Sig}$ & $N_{\rm MC,Sig}$ \\ \hline
$\tautoenunu$& 0.322 $\pm$ 0.040 & 284 $\pm$ 17& 91 $\pm$ 13 & 98 $\pm$ 11 \\
$\tautomununu$		& 0.128 $\pm$ 0.012 & 1070 $\pm$ 33	& 137 $\pm$ 13	&136 $\pm$ 12 	\\
$\tautopinu$		& 0.033 $\pm$ 0.003 & 6990 $\pm$ 80	& 233 $\pm$ 19	&212 $\pm$ 17 	\\
$\tautopipiznu$		& 0.035 $\pm$ 0.005 & 1684 $\pm$ 41	& 59 $\pm$ 9	&62 $\pm$ 9 	\\ \hline
$\bmun$                 & 1.1 $\pm$ 0.6 & 14.0 $\pm$ 3.7	& 15 $\pm$ 10	& 12 $\pm$ 5 \\ \hline
$\ben$			& 0.57 $\pm$ 0.25 & 42 $\pm$ 6	& 24 $\pm$ 11	& 15 $\pm$ 5 \\
\hline
\end{tabular}
\end{center}
\end{table}

\section{SYSTEMATIC STUDIES}
\label{sec:Systematics}

The branching fraction for any of the decay modes in this analysis is given by 

\begin{equation}
\BR (\bln) = \frac{N_{\rm obs} - N_{\rm BG}}{N_{\BB} \eps_{\rm tag} \eps_{\rm sig}},
\end{equation} 

\noindent where $N_{\rm obs}$ is the total number of events observed in the signal region, $N_{\rm BG}$ is the predicted number of events from background in the signal region; we use the values given as $N_{\mbox{\scriptsize{exp,Sig}}}$ in Table \ref{tab:BGpred_EEx} for $\NBG$.  By definition $\Nobs = \Nsig + \NBG$.
$N_{\BB}$ is the total number of $\Y4S$ decays in the data set,  and the efficiencies can have different values for each mode.  Each of these variables, except $N_{\rm obs}$, brings a systematic error into the branching fraction.  

\subsection{Double Tag Control Sample}
\label{sec:DT}

To assess the agreement between Data and MC samples, we used two sets of control samples: sidebands and ``double tagged'' events.  Events where both of the $B$ mesons are reconstructed in tagging modes, 
$B^{-} \rightarrow D^{(*)0} \ell^{-} \overline{\nu_{\ell}}$ vs. 
$B^{+} \rightarrow \overline{D}^{(*)0} \ell^{+} \nu_{\ell}$, are
referred to as ``double tag'' events. Due to both the large branching 
fraction of $D\ell\nu X \xspace$ decays and the 
high tagging efficiency for reconstructing 
these events, a sizable sample ($\approx 3.4 \times 10^{6}$) of such events are 
available in the on-resonance dataset. 

\begin{figure} 
\subfigure[] {\begin{overpic} [width = \twowidefig \linewidth,keepaspectratio]{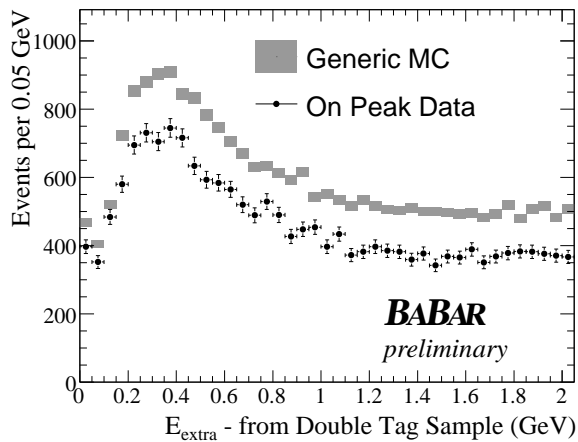}
	\put(63,20){\begin{minipage}{0.6in}\Large
{\bf\babar}\linebreak\small{\em preliminary}
          \end{minipage}
        }
      \end{overpic}
\label{fig:DTEextraStandardCuts}} 
\subfigure[]
{\begin{overpic} [width = \twowidefig \linewidth,keepaspectratio]{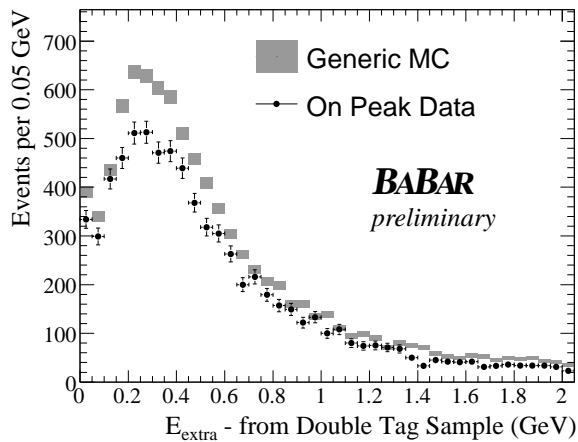}\put(61,42){\begin{minipage}{0.6in}\Large
{\bf\babar}\linebreak\small{\em preliminary}
          \end{minipage}
        }
      \end{overpic}
\label{fig:DTEextra1456Cuts}} 
\subfigure[]
{\begin{overpic} [width = \twowidefig \linewidth,keepaspectratio]{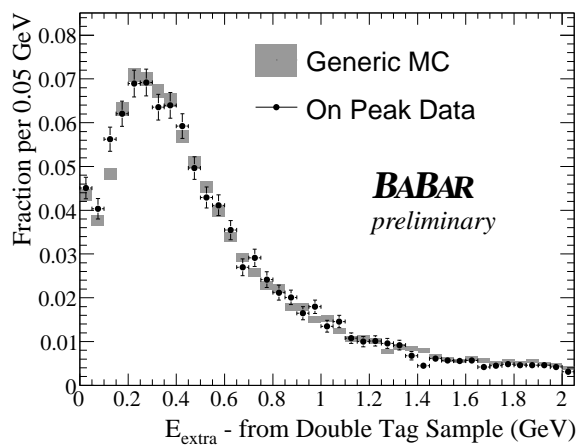}\put(61,42){\begin{minipage}{0.6in}\Large
{\bf\babar}\linebreak\small{\em preliminary}
          \end{minipage}
        }
      \end{overpic}
\label{fig:DTEextra1456CutsUnit}} 
\caption{Distribution of $\eextra$ in double tags with different sets of selection requirements.  The Data and  MC samples are normalized to Data luminosity in \ref{fig:DTEextraStandardCuts} and \ref{fig:DTEextra1456Cuts}.  In \ref{fig:DTEextra1456Cuts}, additional analysis requirements are applied in order to improve agreement.
In \ref{fig:DTEextra1456CutsUnit} the same requirements are applied as in \ref{fig:DTEextra1456Cuts}, but Data and MC are normalized to unit area. The gray rectangles represent the extent of the error bars on the MC histogram.} 
\label{fig:DTEextra} 
\end{figure}

For double tag events, we first applied the tag selection requirements described in section 
\ref{sec:TagReco} to both of the tag candidates.  This procedure resulted in noticeable shape and yield discrepancies between data and MC, as seen in Figure \ref{fig:DTEextraStandardCuts}.  In order to improve agreement, several additional selection criteria were imposed.  These are based on the selection criteria from the previous search for $\btn$ \cite{babar_sl_btn}.  We require that $-2.0 < \cos \theta_{B-D^{0}\ell} < 1.1$ for both $B$ decays and the event has zero net charge.  The resulting distribution is shown in Figure \ref{fig:DTEextra1456Cuts}.  The yield disagreement has improved, and the agreement in shape is better.  To test how much of the disagreement is due to yield and how much is due to shape, we normalize data and MC to unit area with the same cuts as in Figure \ref{fig:DTEextra1456Cuts}.  The resulting distribution is shown in Figure \ref{fig:DTEextra1456CutsUnit}, which shows excellent agreement in the shape of the distribution.

\subsection{\boldmath Systematic Error from Background Prediction ($\NBG$)}
\label{sec:systematicsBG}

We use the ratio of data to MC samples in our background prediction, and the statistical error on that prediction is already large.  The shape of our data and MC samples agree well in the sideband region, so we have no need to apply an additional systematic error.  Therefore, we accept this as the total error and apply no further systematic correction or uncertainty.

\subsection{\boldmath Systematic Error from B-Counting ($N_{\BB}$)}

The estimation of the number of $B^{+}B^{-}$ events present in our data sample has a small uncertainty of 1.1\% \cite{Aubert:2002hc}.

\subsection{\boldmath Systematic Error from Tagging Efficiency $(\eps_{\rm tag}$)}
\label{sec:systematicsEffTag}

We pursue a procedure that attempts to combine what we
know about both the single- and double-tagged samples.  We define the efficiency
\[
\varepsilon_{2}=N_{2}/N_{1},
\] where $N_{1}$ is the single-tagged events, and $N_{2}$ is the number of double-tagged events, as defined in Section \ref{sec:DT}.  We use the ratio of $\varepsilon_{2}$ in Data and our MC samples as a systematic correction to the tag efficiency (0.891).  We take the uncertainty on this ratio as a systematic uncertainty (0.021).  These numbers were extracted using only those events reconstructed as $D^{0} \to K^{-}\pi^{+}$ for the first tag.  We verify that this correction is consistent with the correction calculated using $D^{0} \to K^{-}\pi^{+}\pi^{-}\pi^{+}$ on the first tag.

\subsection{\boldmath Systematic Error from Signal Efficiency $(\eps_{\rm sig}$)}

The systematic uncertainties on the signal selection efficiency for each signal mode have contributions from $\eextra$ mis-modeling, tracking efficiency, particle identification, and $\pi^{0}$ reconstruction for the $\tautopipiznu$ mode.

The systematic correction and uncertainty on the $\eextra$ shape is taken from the double tag $\eextra$ distribution described in Section \ref{sec:DT}.  The double tagged events provide us with a means of comparing data
and simulation, using an independent control sample, to extract this uncertainty.
We use the $\eextra$ distributions shown in Figure \ref{fig:DTEextra1456Cuts} to extract the yield of candidates satisfying $\eextra \leq$ 0.6 GeV. This yield is then compared to the number of candidates in the full distribution.
Comparing the ratio extracted from MC to that extracted from data yields a correction factor, the error on which is taken as the systematic uncertainty for $\eextra$. We extract a correction and uncertainty of $1.015 \pm 0.021$.

Since the particle identification algorithms have not changed since our previous search for $\btn$ \cite{babar_sl_btn}, we use the same values as conservative estimates of our current systematic errors. Since the previous analysis only used the four single-prong $\tau$ decay modes, we must extrapolate.  For the $\tautothreepinu$ mode, we apply the $\pi$ correction three times and take triple the uncertainty as our $a_{1}$ uncertainty.  For the $\bmun$ and $\ben$ modes, we use the $\mu$ and $e$ systematics, which is a conservative choice because particle identification is more effective at the high momenta that characterize two-body $B$ decays. 

All multiplicative contributions to the systematic uncertainty are summarized in Table \ref{tab:systematicCorrErr}.  The corrected efficiencies are shown in Table \ref{tab:corrected_efficiencies}.  We extract a total multiplicative systematic uncertainty of 3.6\% for $\btn$, 4.4\% for $\bmun$, and 4.0\% for $\ben$.

\begin{table}[htb]
\caption{\label{tab:systematicCorrErr} Summary of Systematic corrections, uncertainties, and fractional uncertainties.  Note that the numbers from the rightmost column of Table \ref{tab:corrected_efficiencies} are included in the rightmost column of this Table.}
\renewcommand{\arraystretch}{1.3}
\begin{center}
\begin{tabular}{|c|c|c|c|}
\hline
Source 		& Applicable Mode(s) & Correction & Fractional Uncertainty (\%)\\
\hline
$B$ Counting 	& All 		& 1.0 & 1.1  \\ \hline
Tag efficiency 	& All 		& 0.891 $\pm$ 0.021  & 2.4  \\ \hline
$\eextra$	& All		& 1.015 $\pm$ 0.021 & 2.1 \\ \hline
$\pi^{0}$ Reconstruction& $\tautopipiznu$ & 0.984 $\pm$ 0.030 & 3.0 \\ \hline
Tracking Efficiency 	& $\tautoenunu$ & 1.0 & 0.36 \\
 			& $\tautomununu$ & 1.0  & 0.36 \\
 			& $\tautopinu$ & 1.0& 0.36 \\ 
 			& $\tautopipiznu$ & 1.0  & 0.36 \\
			& $\bmun$ & 1.0  & 0.36 \\	    
			& $\ben$ & 1.0 & 0.36 \\ \hline
Particle Identification	& $\tautoenunu$ & 1.01  & 2.5\\
			&  $\tautomununu$ & 0.92  & 3.1 \\
			&  $\tautopinu$& 1.02 & 0.8 \\
			&  $\tautopipiznu$& 1.00 & 1.5 \\
			&  $\bmun$& 0.92 & 3.1 \\
			&  $\ben$ & 1.01 & 2.5 \\
\hline
\end{tabular}
\end{center}
\end{table}

\begin{table}[htb]
\caption{\label{tab:corrected_efficiencies}%
The corrected tag and signal efficiencies. Two errors are quoted:
the first is the MC statistical uncertainty, and the second is the 
systematic error computed from the sources in this section.}
\renewcommand{\arraystretch}{1.3}
\begin{center}
\begin{tabular}{|c|c|c|}
\hline
Efficiency	&	Corrected								&	Fractional Systematic Error (\%) \\ \hline			
Tag$(\btn)$	&	$(	1.514	\pm	0.003	\pm	0.036	)	\% $	&	2.4	\\		
Tag$(\bmun)$	&	$(	0.937	\pm	0.003	\pm	0.022	)	\% $	&	2.4	\\	
Tag$(\ben)$	&	$(	0.974	\pm	0.003	\pm	0.023	)	\% $	&	2.4	\\	\hline	
$\varepsilon_{\mathrm{sig}}^{(\tautoe)}$	&	$(	2.04	\pm	0.04	\pm	0.07	)	\% $	&	3.3	\\	
$\varepsilon_{\mathrm{sig}}^{(\tautomu)}$	&	$(	1.50	\pm	0.04	\pm	0.06	)	\% $	&	3.7	\\		
$\varepsilon_{\mathrm{sig}}^{(\tautopi)}$	&	$(	2.57	\pm	0.05	\pm	0.06	)	\% $	&	2.2	\\		
$\varepsilon_{\mathrm{sig}}^{(\tautopipiznu)}$	&	$(	0.86	\pm	0.03	\pm	0.03	)	\% $	&	4.0	\\	\hline	
$\varepsilon_{\mathrm{sig}}^{(\btn)}$	&	$(	6.97	\pm	0.08	\pm	0.20	)	\% $	&	2.8	\\	\hline	
$\varepsilon_{\mathrm{sig}}^{(\bmun)}$	&	$(	28.9	\pm	0.3	\pm	1.1	)	\% $	&	3.7	\\	\hline	
$\varepsilon_{\mathrm{sig}}^{(\ben)}$	&	$(	37.9	\pm	0.4	\pm	1.2	)	\% $	&	3.3	\\	\hline	
\end{tabular}
\end{center}
\end{table}

\section{RESULTS}
\label{sec:Physics}

\begin{figure}[htb]
\begin{center}
\hspace{-0.1in}
     \subfigure[]{\begin{overpic}[width=\twowidefig \textwidth]{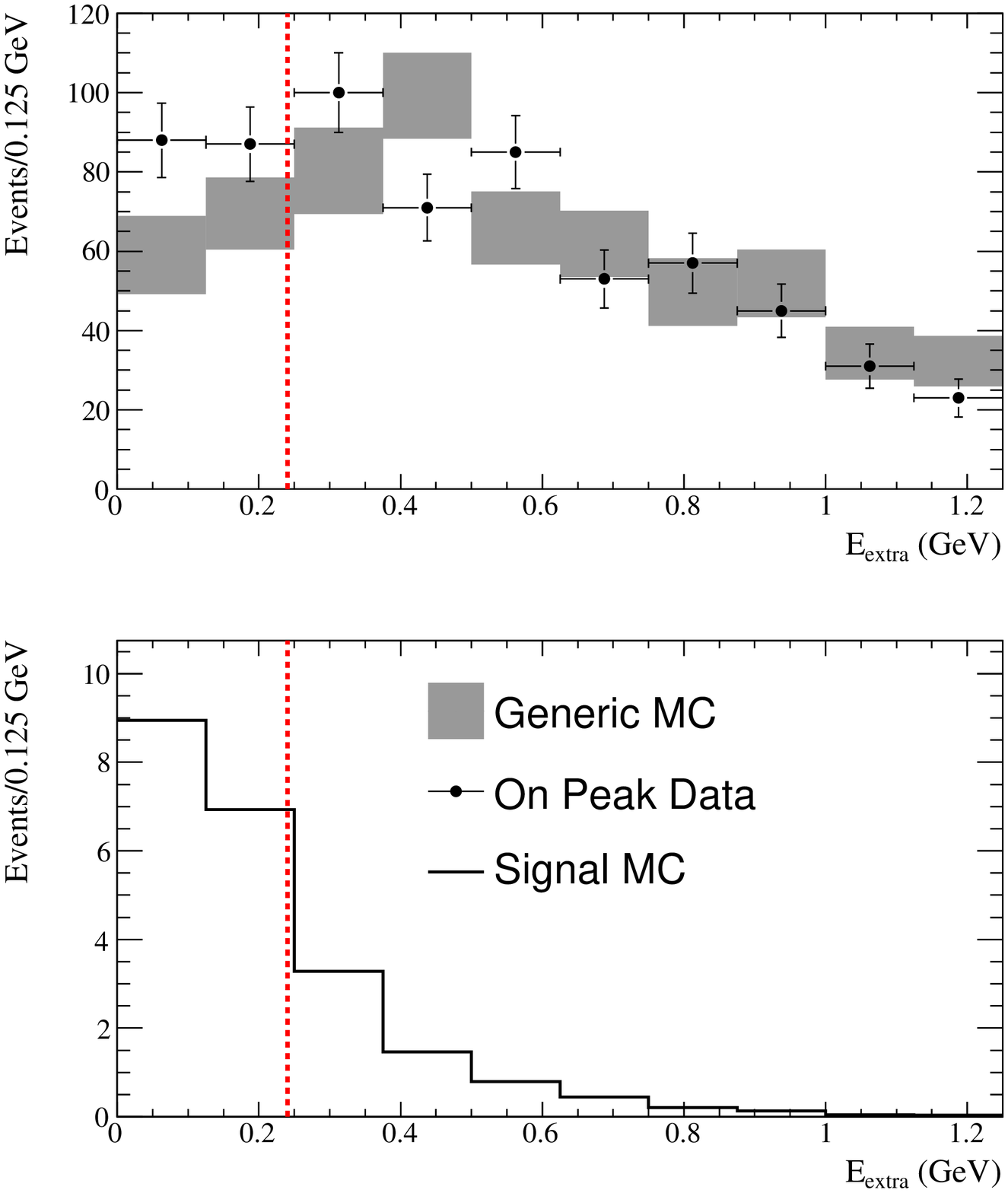}
         \put(60,85){\begin{minipage}{1in}\Large 
{\bf\babar}\linebreak\small{\em preliminary}
          \end{minipage}
        } 
      \end{overpic}
}
     \hspace{.02in}
      \subfigure[]{\begin{overpic} [width=\twowidefig \textwidth]{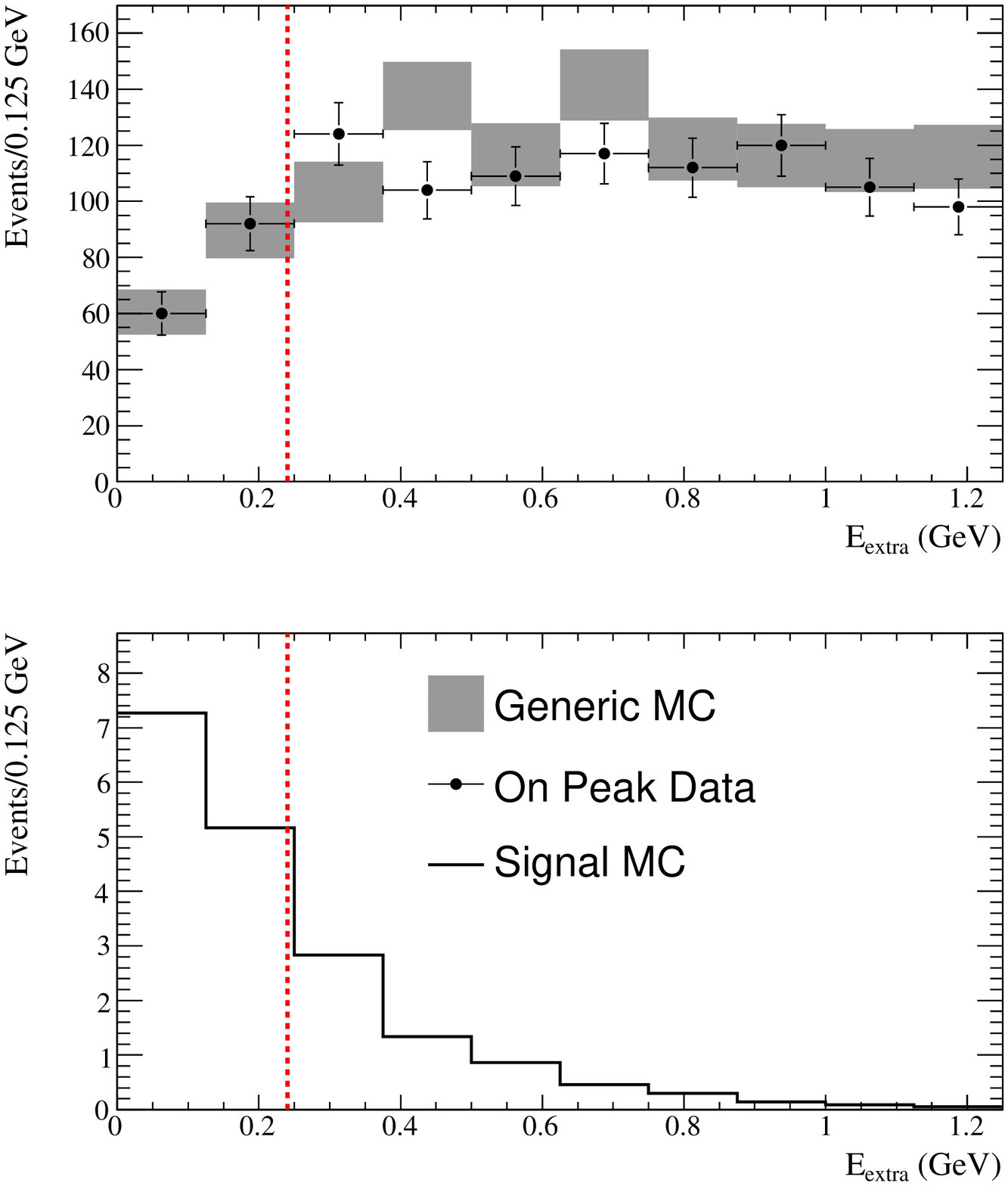}
 \put(40,70){\begin{minipage}{1in}\Large 
{\bf\babar}\linebreak\small{\em preliminary}
          \end{minipage}
        } 
      \end{overpic}
}
\end{center}
\caption{Total extra energy is plotted after all cuts have been applied in 
          the mode (a)~\tautoe~and (b) \tautomu.  The background MC samples have been scaled according
to the ratio of predicted backgrounds from data and MC as presented in section~\ref{sec:EextraSBExtrapolation} and summed together.  The grey rectangles represent the extent of the error bars on the MC histogram.  The signal region consists of the region of $\eextra$ to the left of the vertical dashed line.
	  Simulated $\btn$ signal MC is plotted (lower) for comparison. 
      }
\label{fig:eextra_allcuts_unblind_e}
\end{figure}
\begin{figure}[htb]
\begin{center}
     \subfigure[]{\begin{overpic}[width=\twowidefig \textwidth]{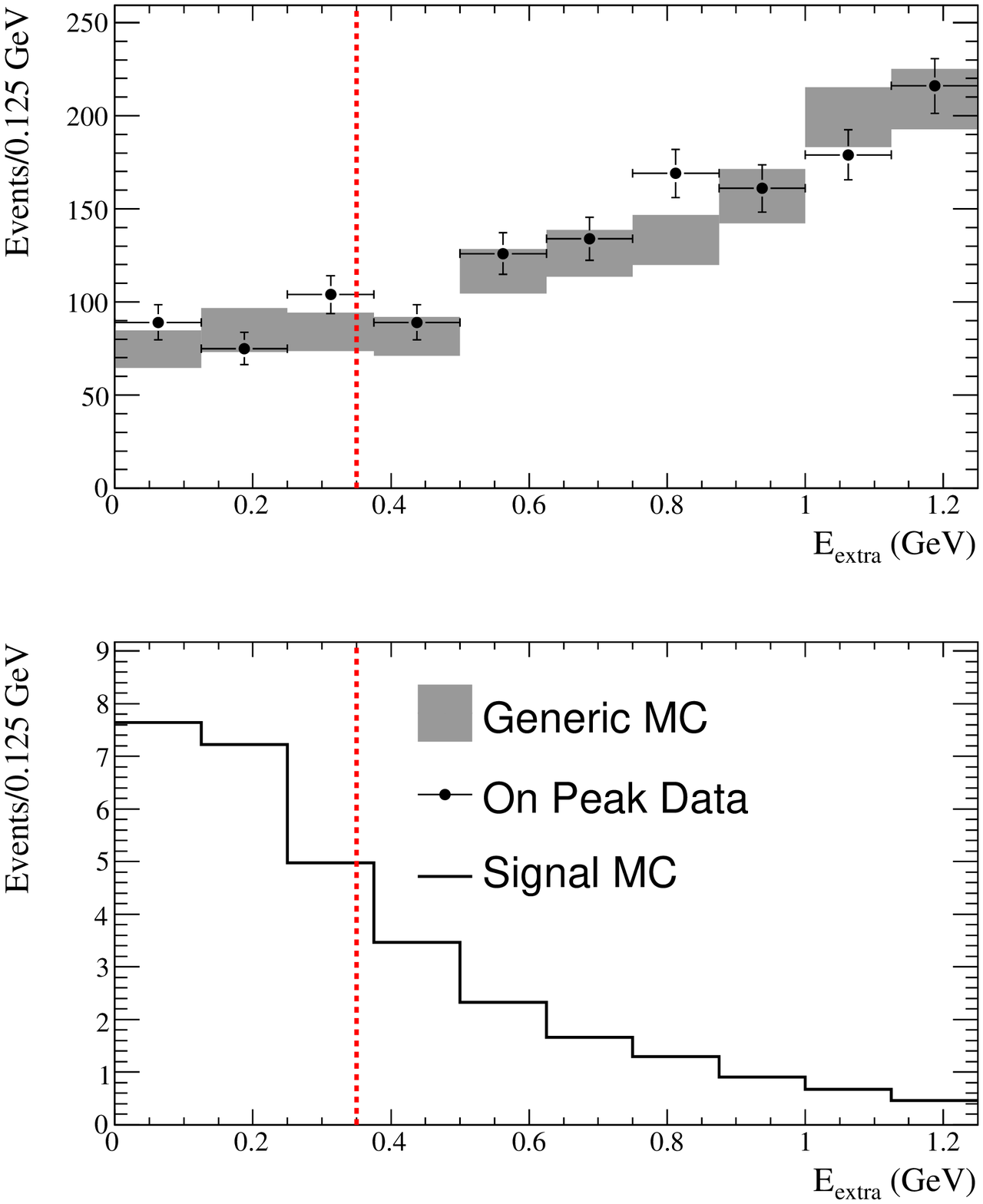}
\put(31,85){\begin{minipage}{1in}\Large 
{\bf\babar}\linebreak\small{\em preliminary}
          \end{minipage}
        } 
      \end{overpic}
}
     \hspace{.02in}
      \subfigure[]{\begin{overpic}[width=\twowidefig \textwidth]{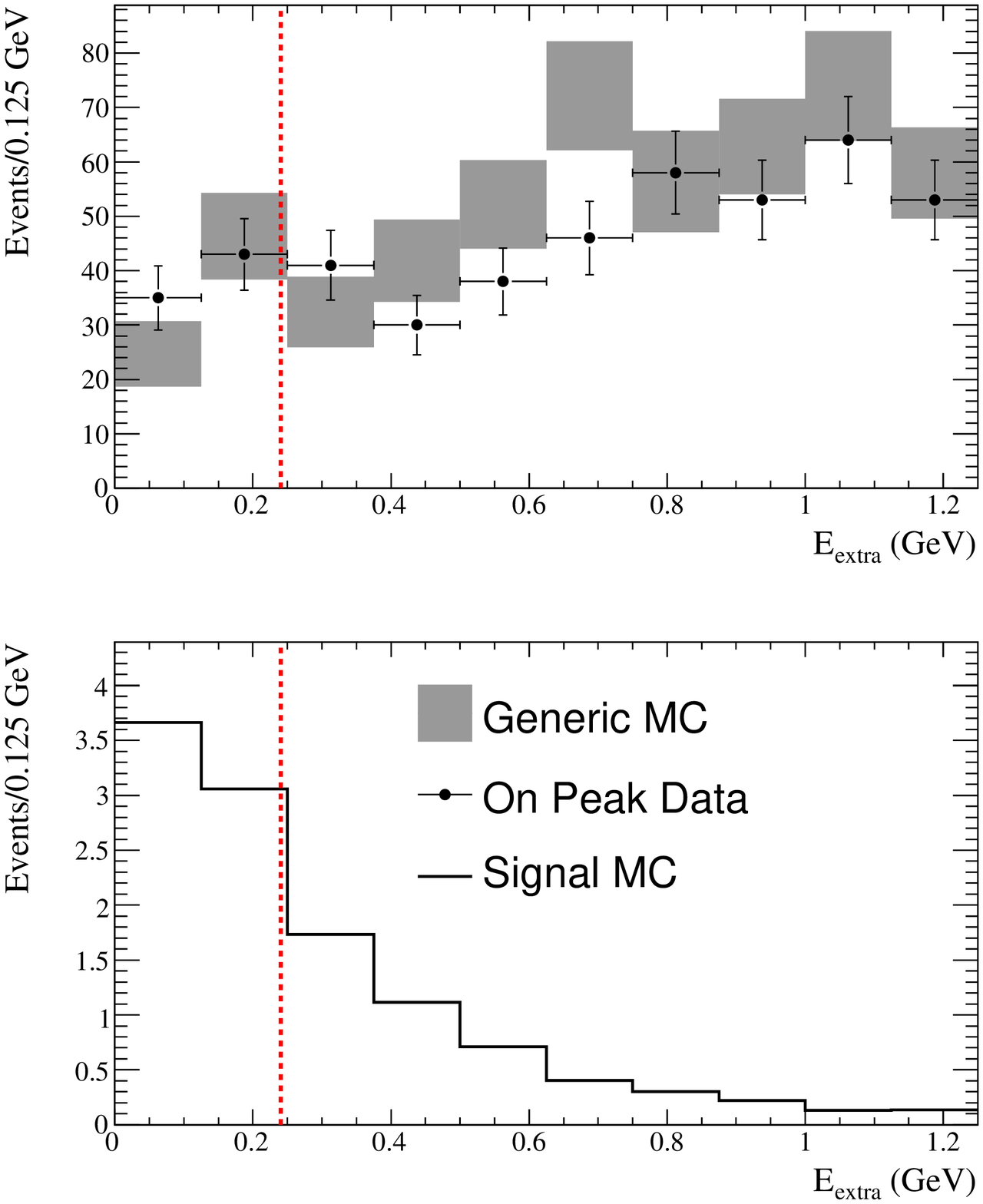}
\put(25,89){\begin{minipage}{1in}\Large 
{\bf\babar}\linebreak\small{\em preliminary}
          \end{minipage}
        }
      \end{overpic}
}
\end{center}
\caption{Total extra energy is plotted after all cuts have been applied in 
          the mode (a) \tautopi~and~(b) \tautorho.  The background MC samples have been scaled according
to the ratio of predicted backgrounds from data and MC as presented in section~\ref{sec:EextraSBExtrapolation} and summed together.  The grey rectangles represent the extent of the error bars on the MC histogram.  The signal region consists of the region of $\eextra$ to the left of the vertical dashed line.
	  Simulated $\btn$ signal MC is plotted (lower) for comparison. 
      }
\label{fig:eextra_allcuts_unblind_mu}
\end{figure}
\begin{figure}[htb]
\begin{center}
     {\begin{overpic}[width=\twowidefig \textwidth]{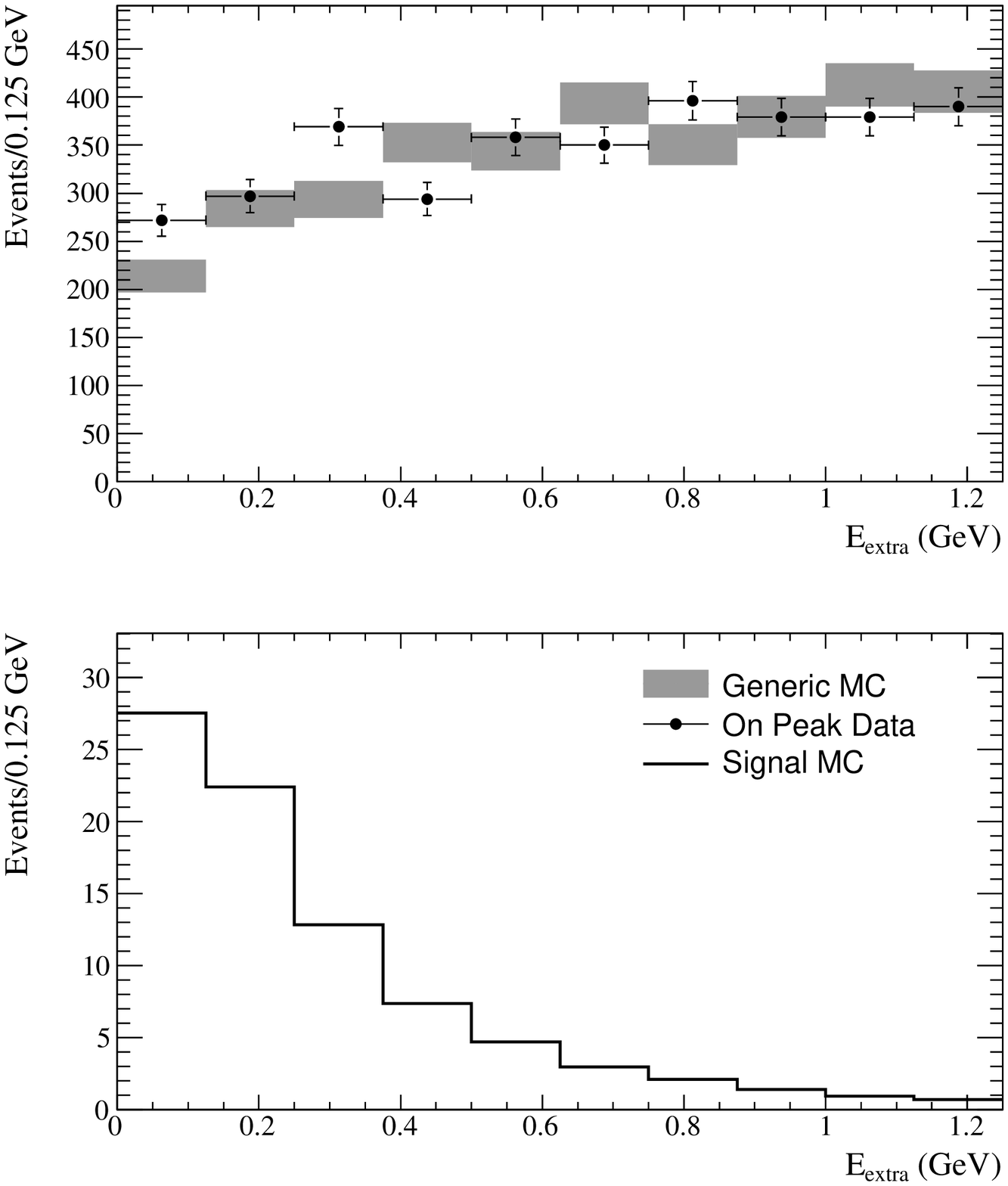}
\put(40,70){\begin{minipage}{1in}\Large 
{\bf\babar}\linebreak\small{\em preliminary}
          \end{minipage}
        }
      \end{overpic}
}
 \end{center}
\caption{Total extra energy is plotted after all cuts have been applied with all $\btn$ modes combined. Events in this distribution are required to pass all selection criteria.  In addition, the background MC samples have been scaled according
to the ratio of predicted backgrounds from data and MC as presented in section~\ref{sec:EextraSBExtrapolation} and summed together.  The grey rectangles represent the extent of the error bars on the MC histogram.
	  Simulated $\btn$ signal MC is plotted (lower) for comparison.
      }
\label{fig:eextra_allcuts_ALL} 
\end{figure}
\begin{figure}[htb]
\begin{center}
     \subfigure[]{\begin{overpic}[width=\twowidefig \textwidth]{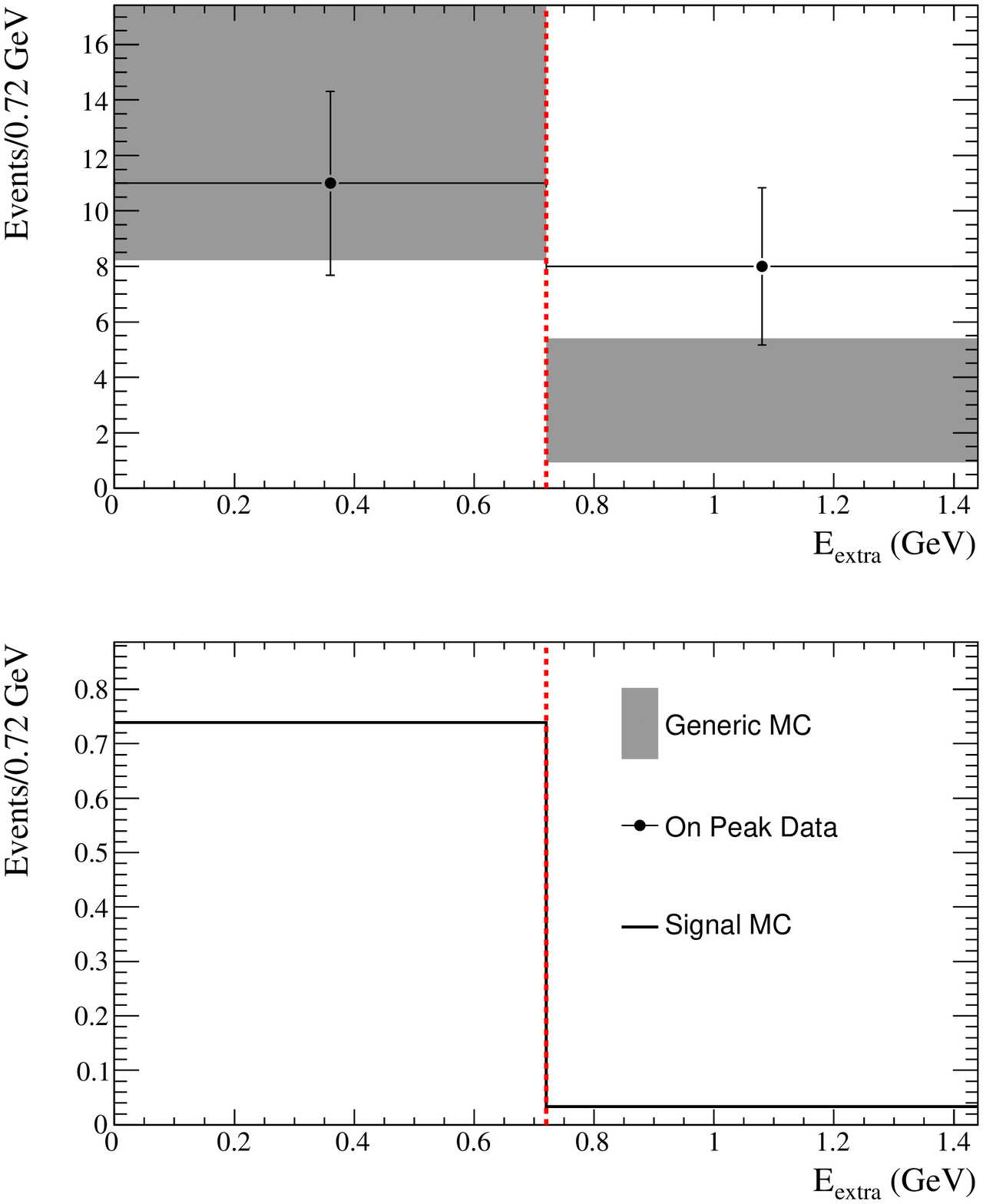}
	\put(60,89){\begin{minipage}{1in}\Large 
	{\bf\babar}\linebreak\small{\em preliminary}
          \end{minipage}
        }
      \end{overpic}
	\label{fig:eextra_allcuts_munu} 
}
     \hspace{.02in}
     \subfigure[]{\begin{overpic}[width=\twowidefig \textwidth]{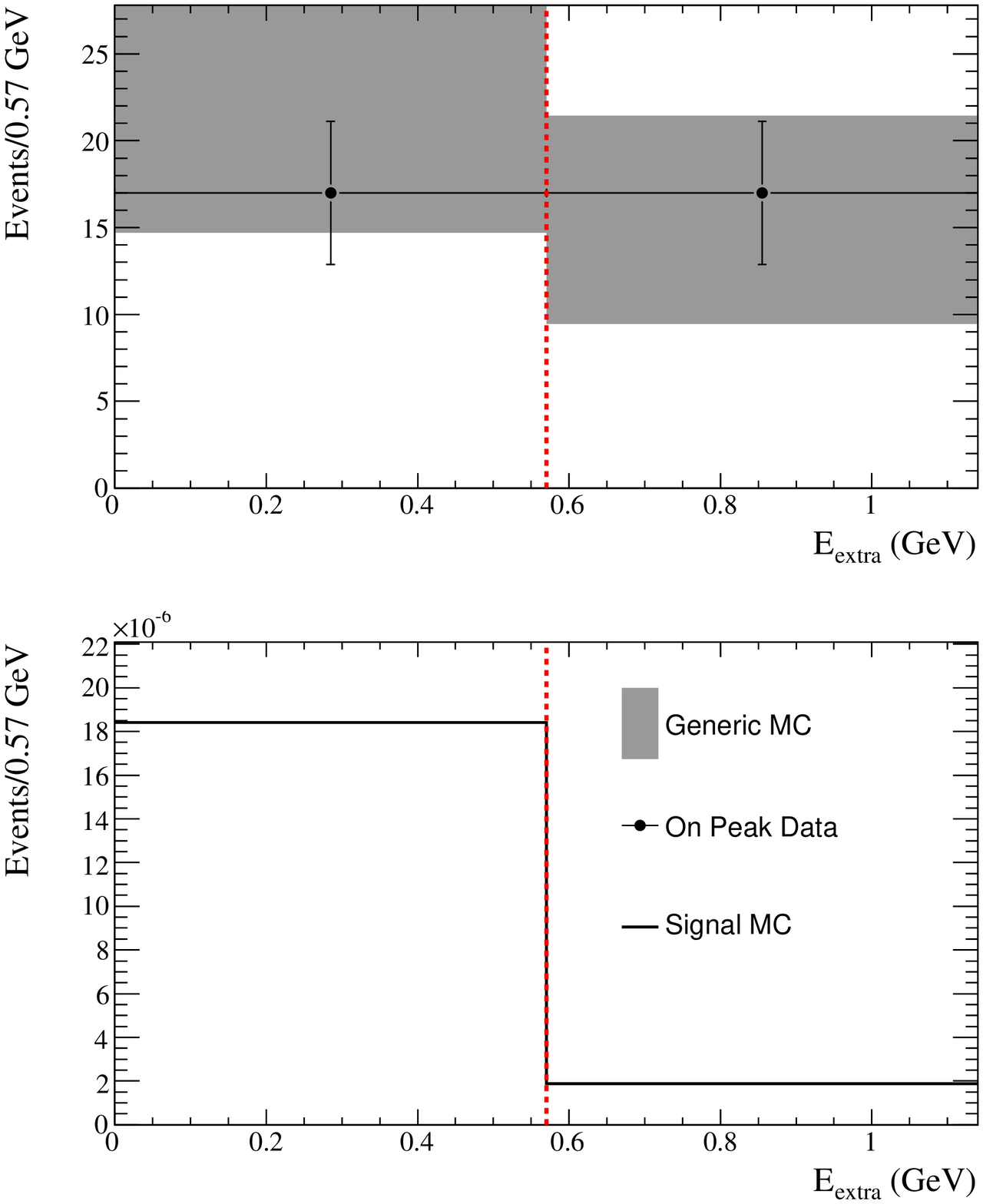}
	\put(50,63){\begin{minipage}{1in}\Large 
	{\bf\babar}\linebreak\small{\em preliminary}
          \end{minipage}
        }
      \end{overpic}
	\label{fig:eextra_allcuts_enu} 
}
\end{center}
\caption{Total extra energy is plotted after all cuts have been applied to $\bmun$(left) and $\ben$ (right). Events in this distribution are required to pass all selection criteria.  The background MC samples have been scaled according to the ratio of predicted backgrounds from data and MC as presented in section~\ref{sec:EextraSBExtrapolation} and summed together.  The grey rectangles represent the extent of the error bars on the MC histogram. The signal region consists of the region of $\eextra$ to the left of the vertical dashed line.
	  Simulated signal MC is plotted (lower) for comparison.
      }

\end{figure}

After finalizing the signal selection criteria, the signal region
in the on-resonance data is examined. 
Table~\ref{tab:Results} lists the
number of observed events in on-resonance data in the signal region,
together with the expected number of background events in the 
signal region. Figures~\ref{fig:eextra_allcuts_unblind_e}~and~\ref{fig:eextra_allcuts_unblind_mu} show the $E_{\rm{extra}}$ 
distribution in data and simulation for each of the $\tau$ decay modes considered.
Data is overlaid on the summed MC contribution, scaled to the dataset luminosity, and signal
MC is plotted for comparison. 
Figure~\ref{fig:eextra_allcuts_ALL} shows the $E_{\rm{extra}}$ distribution for all $\tau$ decay modes combined, with MC scaled to the sideband data yield.  Figure \ref{fig:eextra_allcuts_munu} shows the $E_{\rm{extra}}$ distribution for $\bmun$ with and without MC scaling to the sideband data yield.  Figure \ref{fig:eextra_allcuts_enu} shows the $E_{\rm{extra}}$ distribution for $\ben$ with and without MC scaling to the sideband data yield.

We use the method developed by Feldman and Cousins \cite{FeldmanCousins}, which is designed to produce an upper limit for null results and a two-sided confidence interval for non-null results. The Feldman Cousins method begins with the construction of a confidence belt, which is a two-dimensional histogram $N_{\mathrm{true}}$ vs. $N_{\mathrm{sig}}$.  $N_{\mathrm{sig}}$ represents the extracted signal yield for an ensemble of experiments for each value of $N_{\mathrm{true}}$.  $N_{\mathrm{true}}$ is the actual number of signal events used as the central value to generate each ensemble of experiments.  We generate these distributions using a random number generator.

For each value of $N_{\mathrm{true}}$, we generate two sets of 100,000 random numbers.  For the first set, a Poisson random number generator is used with the central value set to $N_{\mathrm{true}}$.  The second set is based on the background predictions from Table \ref{tab:BGpred_EEx}.  A Gaussian random number generator is used with the center set to the central value from the table (\NBG) and the width equal to $\sqrt{\NBG}$.  This value is then used as the center of another random number distribution ($N_{\mathrm{back}}$), with a width equal to the error on (\NBG) from Table \ref{tab:BGpred_EEx}.  This procedure is used to account for both the error on the background prediction and the statistical error on the total number of observed events ($N_{\mathrm{obs}}$).   These two numbers are summed for each experiment to form $N_{\mathrm{obs}}$.  

\begin{equation}
 N_{\mathrm{sig}} = N_{\mathrm{obs}} - N_{\mathrm{back}}.
\end{equation}

\begin{figure}[htb]
\begin{center}
\hspace{-0.1in}
\subfigure[]{ \begin{overpic}[width=.48\linewidth,keepaspectratio]{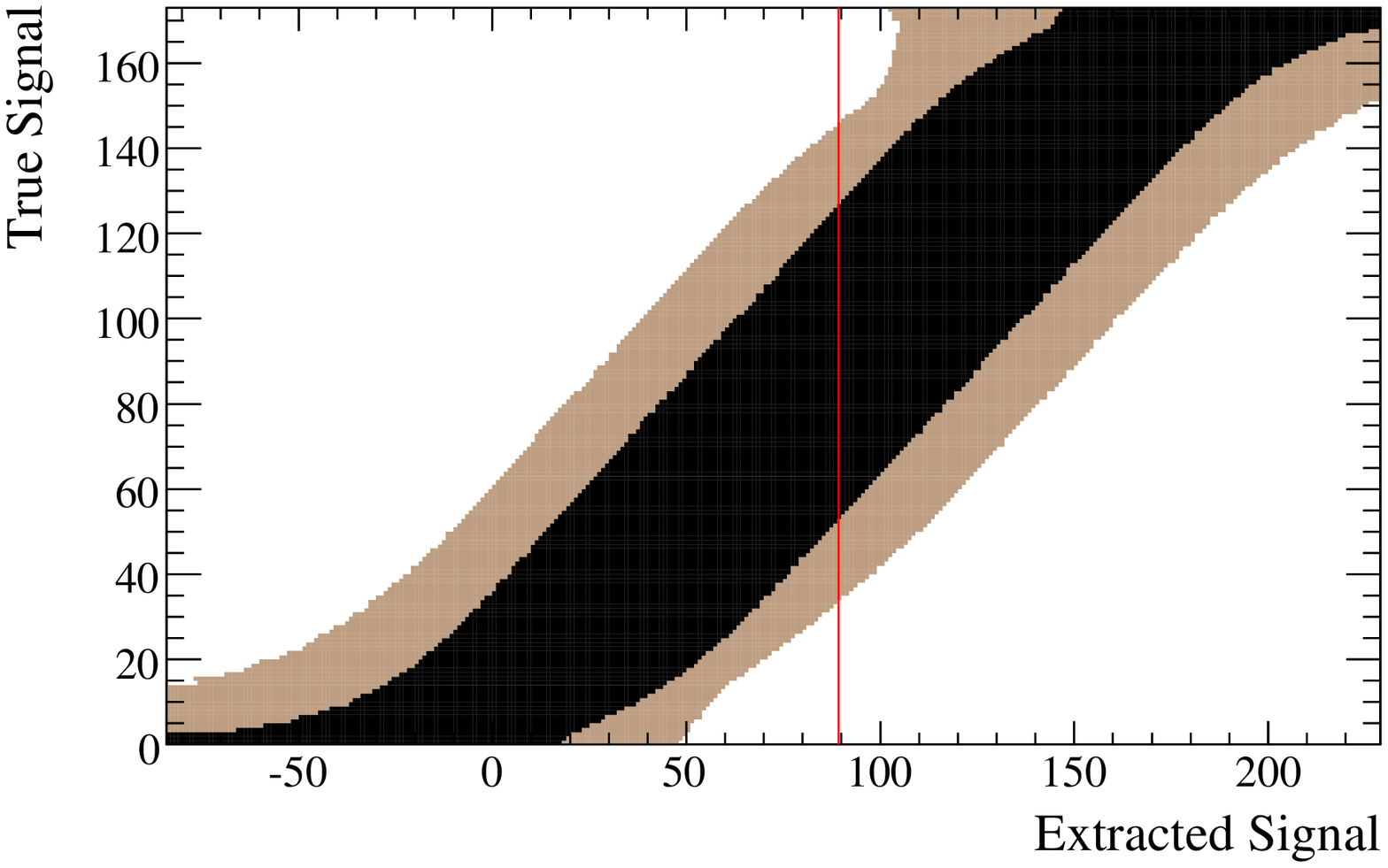}
        \put(15,45){\begin{minipage}{1in}\Large 
{\bf\babar}\linebreak\small{\em preliminary}
          \end{minipage}
        }
      \end{overpic}
}
\subfigure[]{ \begin{overpic}[width=.48\linewidth,keepaspectratio]{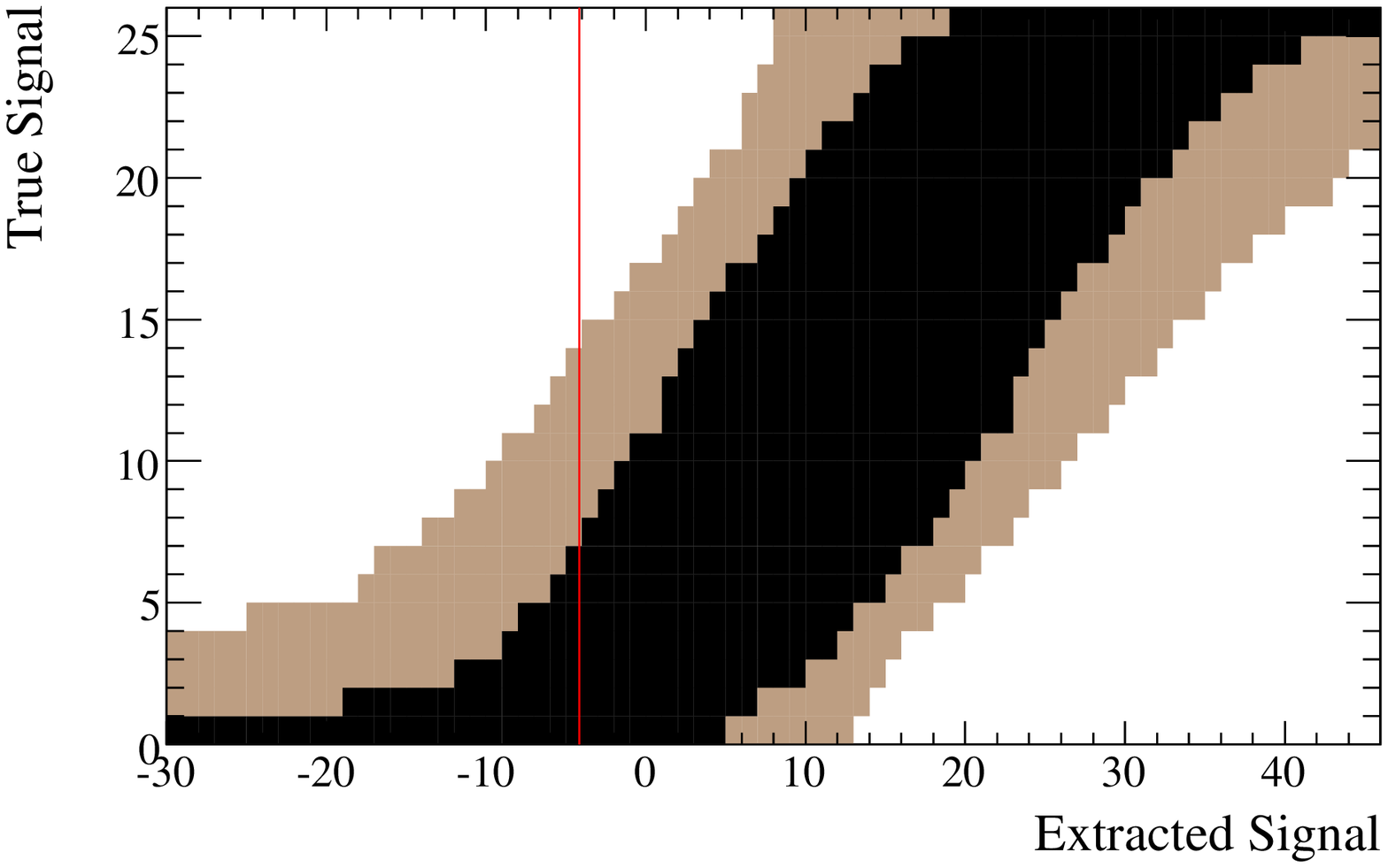}
        \put(15,45){\begin{minipage}{1in}\Large
{\bf\babar}\linebreak\small{\em preliminary}
          \end{minipage}
        }
      \end{overpic}
}
\subfigure[]{ \begin{overpic}[width=.48\linewidth,keepaspectratio]{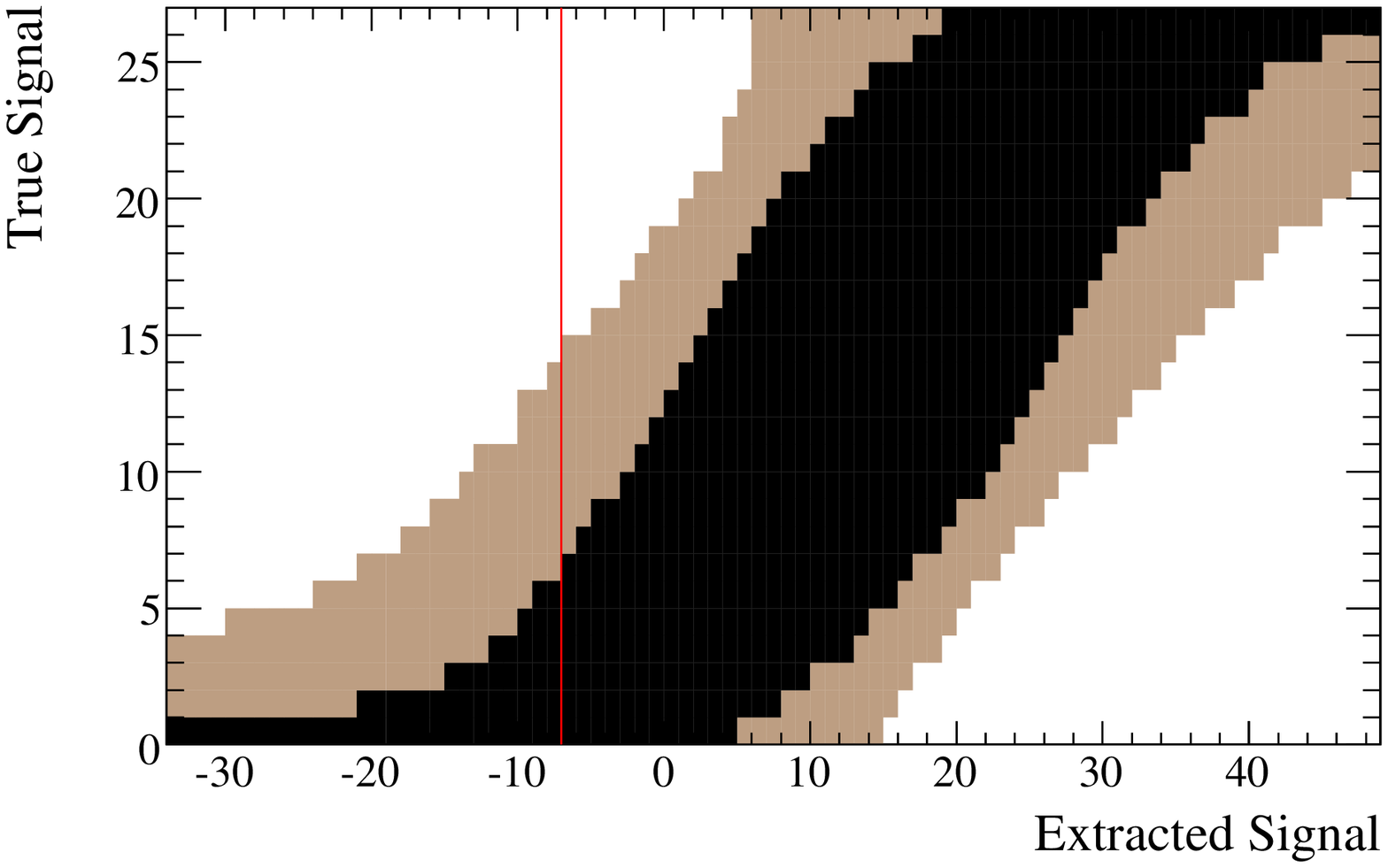}
        \put(15,45){\begin{minipage}{1in}\Large
{\bf\babar}\linebreak\small{\em preliminary}
          \end{minipage}
        }
      \end{overpic}
}
\end{center}
\caption{The confidence band produced by the Feldman Cousins method. The central band represents the $1\sigma$ confidence level; the next band out represents the 90\% confidence level. The vertical red line represents the value of $N_{\mathrm{sig}}$ based on the number of observed events.  Figure (a) is for $\btn$; (b) is for $\bmun$, and (c) is for $\ben$.}
\label{fig:FCCL}
\end{figure}

To smooth the statistical fluctuations resulting from the random number generation, we fit the distribution for each value of $N_{\mathrm{true}}$ to the sum of two Gaussians.  $N_{\mathrm{sig}}$ is then reassigned its value from the fitting functions. For each $N_{\mathrm{true}}$, we must define an acceptance region in $N_{\mathrm{sig}}$ that will determine our upper limit or central value with uncertainty.  The Feldman Cousins method defines this acceptance region without referring to data or any bias regarding whether we seek an upper limit or branching fraction.  For each bin, we calculated the ratio 
\begin{equation}
R \equiv \frac
{P(N_{\mathrm{sig}}|N_{\mathrm{true}})}
{P(N_{\mathrm{sig}}|N_{\mathrm{best}})},
\end{equation}
where $N_{\mathrm{best}}$ is the value of $N_{\mathrm{true}}$ that maximizes the probability of observing $N_{\mathrm{sig}}$.  Thus, $R$ ranges between 0 and 1. 
For each value of $N_{\mathrm{true}}$, we sort the bins of $N_{\mathrm{sig}}$ in order of descending $R$.  The probabilities $P(N_{\mathrm{sig}}|N_{\mathrm{true}})$ are summed in this order until the desired confidence level is reached.  The resulting distributions are shown in Figure \ref{fig:FCCL}.  We calculate $N_{\mathrm{sig}}$ and draw a vertical line through the appropriate distribution at $N_{\mathrm{sig}}$.  The upper and lower limits are determined by the intersection of that line with the appropriate confidence bands.

\begin{table}
\caption{The observed number of on-resonance Data events (\Nobs) in the signal region is shown, together with the number of expected background events (\NBG), the corrected overall selection efficiency ($\varepsilon)$, and the branching fraction calculated from each channel. For the four $\tau$ decay channels, the branching fraction shown is the branching fraction of $\btn$ using only this channel.  All systematic uncertainties are included.
\label{tab:Results}}
\begin{center}
\begin{tabular}{|c|c|c|c|c|} \hline
Mode	& Expected  	& Observed 	& Overall 	& Branching      \\ 
	& Background	& Events	& Efficiency $(\varepsilon)$	& Fraction 	\\
	& (\NBG)	& (\Nobs)	&  & 		\\ \hline
$\tautoenunu$	&	91	$\pm$	13 &	148 & $(	3.08	\pm	0.14	)\times 10^{-4}$ & \btntautoeresult	\\
$\tautomununu$	&	137	$\pm$	13 &	148 & $(	2.28	\pm	0.11	)\times 10^{-4}$& \btntautomuresult	\\
$\tautopinu$	&	233	$\pm$	19 &	243 & $(	3.89	\pm	0.15	)\times 10^{-4}$ & \btntautopiresult	\\
$\tautopipiznu$	&	59	$\pm$	9  &	71  & $(	1.30	\pm	0.07	)\times 10^{-4}$  & \btntautorhoresult \\ \hline
$\btn$		&	521	$\pm$	31 &	610 & $(	10.54	\pm	0.41	)\times 10^{-4}$ & \btnresult	\\ \hline
$\bmun$	        &	15	$\pm$	10  &  11  & $(	27.1	\pm	1.2	)\times 10^{-4}$ & $< \bmunlimit$ @ 90\% CL\\ \hline
$\ben$		&	24	$\pm$	11 &	17  & $(	36.9	\pm	1.5	)\times 10^{-4}$ & $<  \benlimit$ @ 90\% CL   \\ \hline
\end{tabular}
\end{center}
\end{table}

\begin{figure}[htb]
\begin{center}
   	{ \begin{overpic}[width=\twowidefig \textwidth]{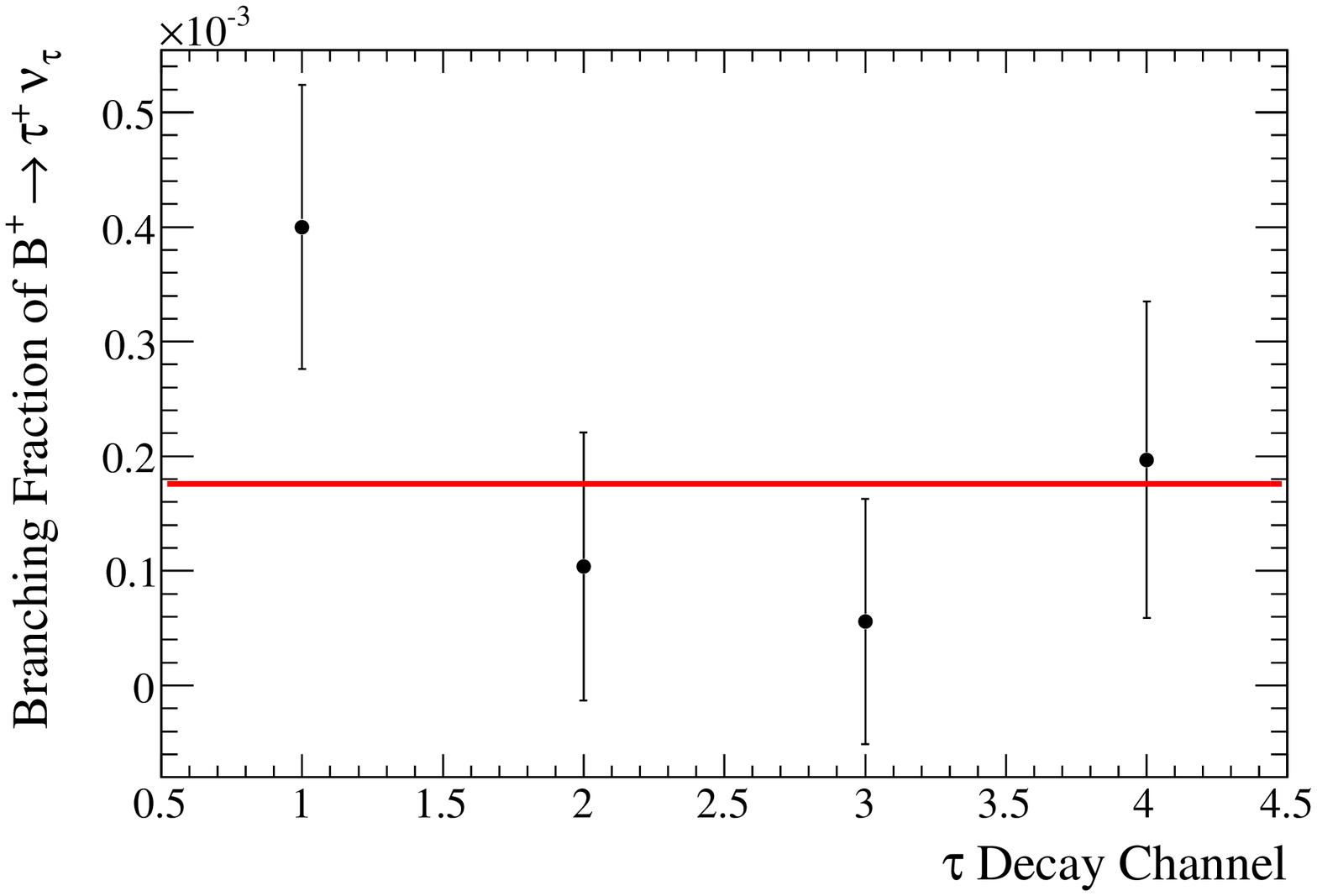}
 \put(55,45){\begin{minipage}{1in}\Large
{\bf\babar}\linebreak\small{\em preliminary}
          \end{minipage}
        }
      \end{overpic}
}
\caption{The branching fractions calculated from each $\tau$ decay channel are plotted.  $1 = \tautoe$, $2 = \tautomu$, $3 = \tautopi$, $4 = \tautorho$.  The horizontal line is a fit to a constant.
      }
\label{fig:BFFits}
\end{center}
\end{figure}

We determine the following preliminary branching fraction:
\begin{equation}
\BR(\btn) = \btnresult.
\end{equation}
Using the Feldman-Cousins method, we determine this result excludes the null hypothesis at the \btnsig\ level, including all systematic uncertainties.

We notice that the electron mode has a much larger excess than the other $\tau$ decay channels.  We performed several tests and cross-checks to determine if this discrepancy is a statistical fluctuation or due to a background mode not included in our MC simulation.  We have tested for the following potential background contributions:   two photon fusion QED events, ``events'' that contain two overlapping $e^{+}e^{-}$ collisions, overzealous Bremsstrahlung recovery, photon pair production where the $e^{+}$ and $e^{-}$ are reconstructed as the tag and signal lepton, and photon pair production events where one lepton is lost and the other is reconstructed as the signal electron.

In two-photon fusion events, the $e^{+}$ and $e^{-}$ each emit a photon; the two photons interact to to produce multiple hadrons.  When these hadrons are misreconstructed as a $D^{0}$ and the $e^{+}$ and $e^{-}$ are identified as a tag and signal lepton, the result is a signal-like event with low extra energy.  Since the signal lepton is an $e^{\pm}$, these events will only populate the electron $\tau$ decay channel.  Two-photon fusion events rarely produce real $D^{0}$ mesons, so they should populate the sidebands and peak of the $D^{0}$ mass distribution.  However, we find that the background predictions taken from he $D^{0}$ mass sidebands (123.8 $\pm$ 17.7 events) is consistent with the prediction from the $\eextra$ sideband (Table \ref{tab:Results}).  In the $\eextra$ distribution for $\tautoenunu$ taken from the sidebands of the $D^{0}$ mass distribution, the data agree will with the MC simulation.  This shows that the excess of signal events in this mode occurs mostly in the peak of the $D^{0}$ mass distribution. Since two-photon fusion events rarely produce real $D^{0}$, any excess from this process should also be in the sidebands. No such excess was present, so we can rule out this potential background. 

We examined the distribution of $\Delta z$, which is the separation between the putative $B$ vertices. Overlapping events should have wider separation than real events, and thus we should see an excess of data above the MC at high values for $\Delta z$. No such excess was found.  Another suggested source of background in this channel is Bremsstrahlung recovery that assigns more photons to an electron than it actually generated.  This could move events into the signal range of $\eextra$ undeservedly.  To test for this, we compare the $\eextra$ distributions for electrons with and without Bremsstrahlung recovery.  When we turn off  Bremsstrahlung recovery, we do so for electrons in both the tag and signal $B$.  The comparison showed that data and MC simulation have very similar shapes regardless of whether Bremsstrahlung recovery is used.  No suspicious excess appears when Bremsstrahlung recovery is activated, so this is not a likely source of the excess.

To remove events in which both daughters of a photon pair conversion are reconstructed as the tag and signal leptons, we require that $m{\ell \ell} > 0.29 \gevcc$ in the $\tautoe$ channel, as described in Section \ref{sec:cutoptimization}.  If a photon produces an $e^{+} e^{-}$ pair in the detector material, one of member of the pair is lost (e.g. down the beam pipe), a $D^{0} \ell^{\pm}$ candidate is reconstructed, and these candidates pass all analysis cuts, we would reconstruct this as a false $\btn$ decay.  However, either member of the $e^{+} e^{-}$ pair is equally likely to be lost, so any excess from this source of background would appear in  events where the tag and signal lepton have the same electrical charge.  However, the MC simulation matches the data for same-sign events very well.

To test the probability of a statistical fluctuation producing the excess seen in the $\tautoe$ channel, we fill a histogram with the branching fraction calculated from each $\tau$ decay channel separately.  The values and uncertainties are taken from the Feldman-Cousins method; systematic uncertainties are not included.  We fit a constant to the branching fractions; the result is $(1.8 \pm 0.6) \times 10^{-4}$.  The fit has a reduced $\chi^{2}$ of 1.64, which corresponds to a probability of 18\%.  The results of this study are shown graphically in Figure \ref{fig:BFFits}.

Using $|V_{ub}| = (4.43 \pm 0.54)\times 10^{-3}$, we extract a preliminary value $f_{B}$ of 
\begin{equation}
	f_{B} = \fB.
\end{equation}

The \babar\ Collaboration previously published a statistically independent measurement of the \btn\ branching fraction using tag $B$ mesons decaying into fully hadronic final states\cite{babar_hadronic_btn}.  We measured
\begin{equation} 
\BR(\btn) = (1.8^{+1.0}_{-0.9}) \times 10^{-4}.
\end{equation}

\noindent Combining these two measurements using a simple error-weighted averaged yields
\begin{equation}
\BR(\btn) = \btnresultcomb,
\end{equation}
\noindent which excludes zero at the $3.2 \sigma$ level.

Since the result for $\BR(\btn)$ is still consistent with zero, we set an upper limit for all three modes at the 90\% confidence level.
\begin{equation}
\BR(\btn) < \btnlimit,
\end{equation}
\begin{equation}
\BR(\bmun) < \bmunlimit,
\end{equation}
\begin{equation}
\BR(\ben) < \benlimit.
\end{equation}

The upper limits for $\BR(\bmun)$ and $\BR(\ben)$ are consistent with previous measurements including the current PDG values of $ < 1.7 \times 10^{-6}$ and $ < 9.8 \times 10^{-7}$, respectively.  The \babar\ Collaboration has set the latest limit on $\BR(\bmun) < 1.3 \times 10^{-6}$ \cite{CONF:inc_bln}.  The limits reported in this analysis are much higher than the best available limits because the tagging method produces a very low background at the cost of a low efficiency.  Low backgrounds are more conducive to discovery; however, until we have enough data to make a statistically significant discovery, other methods will produce stricter upper limits.

\section{CONCLUSIONS}
\label{sec:Summary}

We have performed a search for the decay
process \bln. To accomplish this, a sample of
semileptonic $B$ decays (\btodordszlnu) has been used to
reconstruct one of the $B$ mesons and the remaining information in
the event is searched for evidence of \bln. 

For $\btn$, we measure a preliminary branching fraction of
\[ 
\BR(\btn) = \btnresult,
\]
\noindent which excludes zero at the \btnsig\ level.

We set preliminary upper limits at the 90\% confidence level of 
\[ 
\BR(\btn) < \btnlimit, 
\] 
\[ 
\BR(\bmun) < \bmunlimit, 
\] 
\[ 
\BR(\ben) < \benlimit. 
\]

\section{ACKNOWLEDGMENTS}
\label{sec:Acknowledgments}

We are grateful for the 
extraordinary contributions of our \pep2\ colleagues in
achieving the excellent luminosity and machine conditions
that have made this work possible.
The success of this project also relies critically on the 
expertise and dedication of the computing organizations that 
support \babar.
The collaborating institutions wish to thank 
SLAC for its support and the kind hospitality extended to them. 
This work is supported by the
US Department of Energy
and National Science Foundation, the
Natural Sciences and Engineering Research Council (Canada),
the Commissariat \`a l'Energie Atomique and
Institut National de Physique Nucl\'eaire et de Physique des Particules
(France), the
Bundesministerium f\"ur Bildung und Forschung and
Deutsche Forschungsgemeinschaft
(Germany), the
Istituto Nazionale di Fisica Nucleare (Italy),
the Foundation for Fundamental Research on Matter (The Netherlands),
the Research Council of Norway, the
Ministry of Education and Science of the Russian Federation, 
Ministerio de Educaci\'on y Ciencia (Spain), and the
Science and Technology Facilities Council (United Kingdom).
Individuals have received support from 
the Marie-Curie IEF program (European Union) and
the A. P. Sloan Foundation.

\clearpage
\pagebreak

\bibliographystyle{apsrev}
\bibliography{paper}

\end{document}